\documentclass{aa}

\usepackage[varg]{txfonts}
\usepackage{graphicx}
\usepackage{natbib}
\usepackage{amsmath}
\usepackage{amssymb}
\usepackage{bm}
\usepackage{CJKutf8}
\usepackage{blindtext}
\usepackage{hyperref}
\usepackage{numprint}
\usepackage{xcolor}
\usepackage{multirow}
\usepackage{tabularx}
\usepackage[export]{adjustbox}
\usepackage[rightcaption]{sidecap}
\usepackage[bold=0.5]{xfakebold}
\usepackage{linenoaa}

\renewcommand{\linenumbers}{}

\hypersetup{
    colorlinks=True,
    linkcolor=blue,
    urlcolor=blue,
    citecolor=blue,
    allcolors=blue
}

\begin{document}

\title{Prospects for radio weak lensing: studies using LOFAR observations in the ELAIS-N1 field}

\author{Jinyi Liu (\begin{CJK*}{UTF8}{gbsn}刘晋弋\end{CJK*})\inst{1}
        \and
        Reinout van Weeren\inst{1}
        \and
        Huub Röttgering\inst{1}
        \and
        Konrad Kuijken\inst{1}
        }

\institute{ Leiden Observatory, Leiden University, Einsteinweg 55, 2333 CC Leiden, the Netherlands\\
            \email{jliu@strw.leidenuniv.nl}
            }

\date{Received XXX / Accepted YYY}

\abstract{
We carry out a shape and weak lensing analysis of Low Frequency Array (LOFAR) radio sources and Hyper Suprime-Cam (HSC) optical sources within the European Large Area Infrared Space Observatory Survey-North 1 (ELAIS-N1) field. Using HSC data alone, we detect a cosmic shear correlation signal at a significance of $\sim$$9\sigma$ over a $\sim$$6.4$ deg$^2$ region. For the radio dataset, we analyse observations from both the LOFAR Two Metre Sky Survey (LoTSS) and the International LOFAR Telescope (ILT). While LoTSS provides the deepest radio imaging of ELAIS-N1 with a central source density of $\sim$2.7 arcmin$^{-2}$, its 6\arcsec\  resolution limits the accuracy of shape measurements. But, using LoTSS-matched HSC sources, we show that accurate radio shape measurements would enable us to measure the amplitude of the shear correlation function at least at $\sim$2$\sigma$ significance. In contrast, ILT observation of the field offers a superior $0.3\arcsec$ resolution. By cross-matching HSC and ILT samples, we measure a position angle correlation of $R_{\cos(2\alpha)} = 0.15 \pm 0.02$. This result highlights ILT's ability to resolve extended and diffuse emission. The current ILT observations lack the required depth for robust weak lensing measurements. To assess the potential of ILT, we use simulated data with increased observation hours. Our analysis indicates that with \numprint{3200} hours of ILT observations or deeper data, and assuming that statistical errors dominate over systematics, a shear correlation could be detected with moderate significance. To achieve this will require precise radio shear measurements and effective mitigation of point spread function (PSF) systematics.
}


\titlerunning{Prospects for radio weak lensing: studies with LOFAR observations in the ELAIS-N1 field}
\authorrunning{J. Liu et al.}

\maketitle

\bibpunct{(}{)}{;}{a}{}{,}

\section{Introduction}
The weak gravitational lensing by large-scale structure, known as cosmic shear, describes the coherent distortions of distance galaxy shapes by the gravitational pull of the intervening matter structures \citep[see e.g.][]{Bartelmann2001}. It has become a principal probe of cosmology as it is sensitive to the growth of the cosmic structure and the amplitude of the matter clustering, thereby providing hints on the physical nature of the dark energy and dark matter.

Recent years have seen significant progress in both the methodologies and the observations for weak lensing measurements particularly through Stage III weak lensing surveys, which include the Dark Energy Survey \citep[DES;][]{TheDarkEnergySurveyCollaboration2005, Abbott2018}, the Kilo Degree Survey \citep[KiDS;][]{deJong2013, Kuijken2015}, and the Hyper Suprime-Cam Subaru Strategic Program \citep[HSC-SSP;][]{Aihara2018a, Hamana2020}. The future Vera C. Rubin Observatory's Legacy Survey of Space and Time \citep[LSST;][]{Ivezic2019} and the currently ongoing \textit{Euclid} mission \citep{Laureijs2011, Amendola2018, EuclidCollaboration2024} also aim to measure weak gravitational lensing with high significance. However, to date, the cosmic shear has only been detected with optical or infrared observations due to their large galaxy number densities; for example, KiDS and DES surveys reach number densities of $\sim$6 $\mathrm{arcmin}^{-2}$ \citep{Gatti2021, Giblin2021}, HSC of $\sim$20 $\mathrm{arcmin}^{-2}$ \citep{Mandelbaum2018Jan}, and the future Euclid of $>$30 $\mathrm{arcmin}^{-2}$.

In the next decade, this will no longer be the case with the advent of the next-generation radio telescopes such as the Square Kilometre Array (SKA)\footnote{\url{https://www.skao.int/}}. The continuum weak lensing survey of SKA Phase 1 intends to achieve a usable galaxy number density of 2.7 $\mathrm{arcmin}^{-2}$ for weak lensing, a resolution of $\sim$0.3 arcsec at 1.4 GHz, and a sky coverage of \numprint{5000} deg$^2$, therefore, making it possible to perform accurate radio weak lensing measurements in the forthcoming future \citep{Brown2015, Harrison2016, Bonaldi2016, Camera2017, SquareKilometreArrayCosmologyScienceWorkingGroup2020}. 

There are several unique advantages of weak lensing measurements in the radio band. The most promising aspect of radio weak lensing is that it offers access to higher redshift galaxies, which are expected to be more sensitive to the weak lensing effect and therefore can help probe a greater range of the cosmic history. Besides, by cross-correlating with optical weak lensing surveys, the systematics correlation can be efficiently mitigated due to the distinct designs of telescopes operating in different bands \citep{Brown2015}. Furthermore, the correlation between the intrinsic alignments can be potentially reduced by including the radio polarisation information \citep{Brown2011}.

However, until now, the only possibly successful detection of the radio cosmic shear auto-correlation signal was carried out by \cite{Chang2004}, where the weak lensing E-mode signal was measured at 3.6$\sigma$ level on angular scales from 1$^\circ$ to 4$^\circ$ using the Faint Images of the Radio Sky at Twenty centimetres survey (FIRST; \citealp[see][]{Becker1995, White1997}). Another tentative detection of the radio cosmic shear used cross-correlation between Sloan Digital Sky Survey (SDSS) optical and FIRST radio galaxy shapes and measured a E-mode shear signal at 2.7$\sigma$ level \citep{Demetroullas2016}. Notably, in a study of radio-optical galaxy-galaxy lensing, \cite{Demetroullas2018} measured significant tangential shear signals using FIRST radio sources as the background sample, and lens samples from SDSS. Attempts have also been made in the Hubble Deep Field North (HDFN) by \cite{Patel2010} with Very Large Array (VLA) and Multi-Element Radio-Linked Interferometer Network (MERLIN) observations, and the Cosmological Evolution Survey (COSMOS) deep field by \cite{Tunbridge2016} and \cite{Hillier2019} using 1 GHz and 3 GHz VLA observations, respectively. However, these studies show no detection of the radio weak lensing signal. Recently, \cite{Harrison2020} presented the weak lensing analysis results for the first data release of the SuperCluster Assisted Shear Survey (SuperCLASS; \citealp{Battye2020}), the first radio survey designed primarily for weak lensing studies. They analysed 0.26 deg$^2$ of enhanced MERLIN (\textit{e}-MERLIN) and Karl G. Jansky VLA (JVLA) radio data, and applied a new shape measurement method for interferometric synthetic image, yet failed to detect radio shear signal in both radio shear auto power spectra and radio-optical shear cross power spectra, mainly due to the low source number densities (0.05 arcmin$^{-2}$ in the \textit{e}-MERLIN data and 0.47 arcmin$^{-2}$ in the JVLA data).

In our work, we evaluate the radio-optical shape correlations and the future possibility of detecting weak lensing signal with the Low Frequency Array (LOFAR; \citealp{vanHaarlem2013}). Over the past years, LOFAR has achieved high-sensitivity radio observations with both Dutch LOFAR stations\citep[e.g.,][]{Tasse2021, Sabater2021, Shimwell2024} and International LOFAR Telescope (ILT; \citealp{Morabito2022, Sweijen2022, deJong2024}). In particular, the sub-arcsecond resolution of ILT enables LOFAR to resolve extended emissions, such as those from star-forming galaxies (SFGs). Exploring radio-optical shape correlations serves as an initial check on the noise levels in radio data from LOFAR and offers insight into the reliability of radio-based weak lensing measurement.

This paper is organized as follows. In Sect.~\ref{sec:theory}, we describe the relevant background theory behind weak gravitational lensing and the shear statistics that we use throughout the analysis. Section.~\ref{sec:data} provide the details on the selection process of the radio and optical samples, along with the methods for the source ellipticity or shear calculations. In Sect.~\ref{sec:hsc_2pcf}, we present the shear correlation results derived from the optical data. Subsequently, Sect.~\ref{sec:compare} compares the radio and optical samples, particularly focusing on their shapes. We provide predictions for weak lensing measurements based on ultra-deep observations with ILT in Sect.\ref{sec:forecast}. Finally, we discuss and summarise our work in Sect.~\ref{sec:discussion} and Sect.~\ref{sec:conclusions}.

\section{Weak lensing theory}
\label{sec:theory}
We provide a concise summary of the basic weak lensing theory in this section and refer the reader to the reviews by \cite{Bartelmann2001} and \cite{Kilbinger2015} for further details.

Gravitational tidal field of large-scale structure deflect the light rays from distant sources and results in a mapping between the image position $\boldsymbol{\theta}$ and source position $\boldsymbol{\beta}$ via the deflection angle $\boldsymbol{\alpha}$, know as the lens equation, $\boldsymbol{\beta}=\boldsymbol{\theta}-\boldsymbol{\alpha}$. The deflection angle $\boldsymbol{\alpha}$ can be written as the gradient of the lensing potential, $\boldsymbol{\alpha}=\nabla \psi$, where the lensing potential $\psi$ is related to the gravitational potential $\Phi$ through the integral
\begin{equation}
    \psi(\boldsymbol{\theta}) = \frac{2}{c^2}\int_0^{\chi_\mathrm{H}}\mathrm{d}\chi\ q(\chi)\Phi[f_K(\chi)\boldsymbol{\theta},\chi] .
\end{equation}
Here, $\psi$ is defined as the projected 2D potential, $\chi$ is the comoving distance, $f_K(\chi)$ is the comoving angular diameter distance, $\chi_\mathrm{H}$ corresponds to the horizon distance, and the weighting function $q(\chi)$, also known as the lensing efficiency, takes the form
\begin{equation}
    q(\chi)=\int_{\chi}^{\chi_\mathrm{H}} \mathrm{d}\chi^\prime\ n(\chi^\prime)\frac{f_K(\chi^\prime-\chi)}{f_K(\chi^\prime)},
\end{equation}
where $n(\chi)$ represents the normalized source distribution of the galaxy sample.

In the weak lensing limit, we can linearise the lens equation and define a Jacobian matrix:
\begin{equation}
    \mathbf{A} = \frac{\partial\boldsymbol{\beta}}{\partial\boldsymbol{\theta}} = \begin{pmatrix}
    1-\kappa-\gamma_1 & -\gamma_2 \\
    -\gamma_2 & 1-\kappa+\gamma_1
    \end{pmatrix},
\end{equation}
where $\kappa$ is the convergence, $\gamma_1$ and $\gamma_2$ are the two shear components. By definition, they are second derivatives of the lensing potential
\begin{equation}\label{eq:kappa_shear}
    \begin{aligned}
        &\kappa=\frac{1}{2}(\partial_1^2+\partial_2^2)\psi=\frac{1}{2}\nabla^2\psi, \\
        &\gamma_1=\frac{1}{2}(\partial_1^2-\partial_2^2)\psi,\ \ \ \gamma_2=\partial_1\partial_2\psi.
    \end{aligned}
\end{equation}
The convergence determines the isotropic change of the source size on the observed image. The shear, being the traceless part of the matrix, quantifies the anisotropic stretching of the source and is the primary observable weak lensing effect which allows us to extract valuable cosmological information.

For the choice of shear statistics, we use the classic shear two-point correlation function (2pCF) since it can be easily measured by multiplying the ellipticities of galaxy pairs. The shear 2pCF is expressed in terms of tangential shear and cross shear components, $\gamma_t$ and $\gamma_\times$, which are defined for each galaxy pair and are decomposed from $\gamma_1$ and $\gamma_2$ parameters:
\begin{equation}
    \begin{aligned}
        &\gamma_t=-\gamma_1\cos2\phi - \gamma_2\sin2\phi, \\
        &\gamma_\times=\gamma_1\sin2\phi - \gamma_2\cos2\phi,
    \end{aligned}
\end{equation}
where $\phi$ is the position angle of the line joining the two galaxies. The shear 2pCF can then be defined as
\begin{equation}\label{eq:2pCF}
\xi_\pm(\theta)=\langle\gamma_t\gamma_t\rangle(\theta)\pm\langle\gamma_\times\gamma_\times\rangle(\theta)=\int\frac{\ell\mathrm{d}\ell}{2\pi}P_\kappa(\ell)J_{0/4}(\ell\theta),
\end{equation}
where a connection to the lensing convergence power spectrum $P_\kappa(\ell)$ is also provided here with $J_{0/4}$ denoting the zeroth-order and fourth-order Bessel functions in the $\xi_+$ and $\xi_-$ integrals. Applying Limber projection, $P_\kappa(\ell)$ is related to the 3D matter power spectrum by \citep{Limber1953}
\begin{equation}
    P_\kappa({\ell})=\frac{9H_0^4\Omega_m^2}{4c^4}\int_0^{\chi_\mathrm{H}}\mathrm{d}\chi\ \frac{q^2(\chi)}{a^2(\chi)} P_\delta\left(\frac{\ell}{f_K(\chi)}, \chi\right),
\end{equation}
which establishes a link from the shear two-point statistics in Eq.~(\ref{eq:2pCF}) to the distribution of matter in the Universe.

\section{Data}
\label{sec:data}

\begin{figure}
    \includegraphics{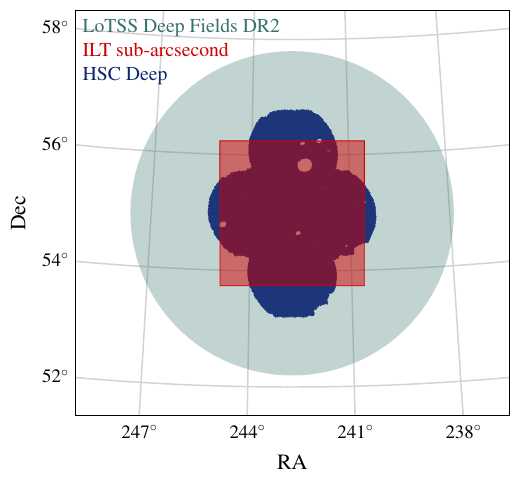}
    \caption{Sky coverage of ELAIS-N1 data from LoTSS Deep Fields DR2 ($\sim$24.5 deg$^2$), ILT sub-arcsecond observation ($\sim$6.7 deg$^2$), and HSC Deep Fields ($\sim$6.4 deg$^2$). The holes in HSC coverage are bright star masks.}
    \label{fig:moc}
\end{figure}

The European Large Area Infrared Space Observatory-North 1 (ELAIS-N1) field was originally chosen as one of the regions in the Northern Hemisphere for the ELAIS survey \citep{Oliver2000} and has since been covered by an extensive range of deep and wide multi-wavelength observations \citep[see][for a detailed description of the multi-wavelength coverage]{Kondapally2021}. The data used in our analysis for the ELAIS-N1 field come from three surveys: the LOFAR Two Metre Sky Survey (LoTSS) Deep Fields, which provides the deepest radio observation of the field to date; the ILT sub-arcsecond observation of ELAIS-N1 field, which provides the radio observation with the exceptional sub-arcsecond resolution; and the HSC-SSP survey, which offers a large sample of optical sources with accurate shape and photometric redshift measurements. The sky coverage of these data is shown in Fig.~\ref{fig:moc}.

\subsection{LoTSS Deep Fields data}
\label{subsec:lofar}
LoTSS is a large radio imaging survey conducted by the LOFAR telescope with the goal of covering the entire northern sky at the frequencies ranging from 120 to 168 MHz \citep{Shimwell2017, Shimwell2019, Shimwell2022}. The LoTSS Deep Fields aims to conduct deep radio imaging over select regions, including the ELAIS-N1 field, and are able to probe fainter and higher redshift radio sources where the dominant populations are SFGs and radio quiet active galactic nucleus (AGN) instead of the radio loud AGN.

The deep-field ELAIS-N1 data we used in this work are from LoTSS Deep Fields Data Release 2 \citep{Shimwell2024}. With a total observing time of $\sim$500 hours using the LOFAR stations within the Netherlands, the image reaches a resolution of $\sim$6$^{\prime\prime}$ and an average noise level of $\lesssim$15 $\mathrm{\mu Jy\ beam^{-1}}$ in the inner 5 deg$^2$ ($\sim$11 $\mathrm{\mu Jy\ beam^{-1}}$ at the centre of the field). Direction-dependent calibration and imaging techniques \citep[see][]{Tasse2021} were employed during the image synthesis process, in order to compensate for the direction-dependent effects (DDEs) such as ionospheric distortions. From the image, the ELAIS-N1 source catalogue is extracted with the Python Blob Detector and Source Finder\footnote{\url{https://pybdsf.readthedocs.io/}} (\textsc{PyBDSF}; \citealp{Mohan2015}) where single or multiple Gaussians are fit to each detected source. The full catalogue contains \numprint{154952} sources, with a mean source density of $\sim$1.8 arcmin$^{-2}$ ($\sim$2.7 arcmin$^{-2}$ in the inner 5 deg$^2$ region).

We do not use the shape information measured from this survey, as the PSF in the LoTSS Deep Fields is too large to permit accurate shape measurements for galaxies. Instead, we utilise the full catalogue to cross-match with the optical dataset and compute the shear correlation from the optical counterparts (see Sect.~\ref{sec:hsc_2pcf}).

\subsection{ILT data}
\label{subsec:ILT}

The radio data obtained in the previous section only makes use of Dutch LOFAR stations, which have baselines up to approximately 120 km. Recently, \cite{deJong2024} released a sub-arcsecond ELAIS-N1 image using all Dutch and international LOFAR stations, extending the maximum baseline to around \numprint{2000} km, and thus providing a significant advantage in terms of angular resolution over Dutch LOFAR.

The ILT sub-arcsecond data of ELAIS-N1 includes four 8-hour observations and achieves a highest resolution of 0.3$^{\prime\prime}$. Due to the increased number of stations in ILT, the RMS noise level reaches 14 $\mathrm{\mu Jy\ beam^{-1}}$ at the centre of the field, comparable to that of a 500-hour observation with Dutch LOFAR (see Sect.~\ref{subsec:lofar}). Similar to the imaging and cataloguing processes in the 500-hour ELAIS-N1 data, the imaging process for ILT data also employs direction-dependent calibration techniques, and \textsc{PyBDSF} is also used for cataloguing. The full catalogue at 0.3$^{\prime\prime}$ resolution contains \numprint{9203} sources and it covers a sky area of $\sim$6.7 deg$^2$.

A series of cuts were applied to the catalogue to reduce the systematics for later analysis. Sources with complex morphologies were removed by setting the source code \texttt{S\_Code="S"}\footnote{This criterion is imperfect for selecting simple morphologies, as some high S/N SFGs are excluded due to internal drawback in the \textsc{PyBDSF} code. See Sect.~\ref{subsec:limitations} for further discussion.}. Additionally, unresolved sources, whose \textsc{PyBDSF} deconvolved major and minor axes (\texttt{DC\_Maj} and \texttt{DC\_Min} columns in the catalogue) are equal to zero, were discarded from the catalogue. To select only the extended sources, we used a simple size cut, \texttt{Maj} $>$ \texttt{Maj\_PSF}, to reject the sources that are too small and thus can be easily influenced by the point spread function (PSF). After applying these cuts, we were left with \numprint{7211} sources, corresponding to $\sim$78\% of the full catalogue. Table~\ref{table:cuts on lofar high res} summarises these cuts. A large fraction of sources are preserved after these cut owing to the exceptional angular resolution attained by international LOFAR.

We will make use of the \textsc{PyBDSF} shapes measured by ILT in Sect.~\ref{subsec:shape_comparison} for radio-optical shape comparison. The ellipticity $\bm{e}$ for each source is calculated from the \textsc{PyBDSF} major axis $a$, minor axis $b$, and the position angle $\alpha$,
\begin{equation}
\label{eq:ellipticity}
    \bm{e}=\frac{a^2-b^2}{a^2+b^2}(\cos2\alpha, \sin2\alpha),
\end{equation}
where we have converted the \textsc{PyBDSF} position angle \texttt{PA}, measured east of north, to the angle measured north of west, which is conventionally used in weak lensing studies. $a$ and $b$ are deconvolved major and minor axes. The uncertainty of the ellipticity, $\sigma_e$, can be estimated by propagating the uncertainties of $a$, $b$, and $\alpha$ through the above equation. Typical measurement uncertainties for $a$, $b$, and $\alpha$ are $\sqrt{\langle\delta a^2\rangle}=0.56$\arcsec, $\sqrt{\langle\delta b^2\rangle}=0.34$\arcsec, and $\sqrt{\langle\delta \alpha^2\rangle}=37.6^\circ$. These combine to yield an effective shape measurement noise of $\sqrt{\langle\sigma_e^2\rangle}=0.236$ per component for the ILT sample.

\begin{table}
\caption{Summary of the cuts applied to the ILT sub-arcsecond ELAIS-N1 catalogue.}
\centering
{\renewcommand{\arraystretch}{1.3}
\label{table:cuts on lofar high res}
\begin{tabular}{l r r}
\hline\hline
Cut & $N_\mathrm{g}$  & Percentage \\
\hline
Full catalogue                                  & \numprint{9203}  & 100\% \\
Simple morphology                               & \numprint{8228}  & 89\% \\
Resolved                                        & \numprint{7669}  & 83\% \\
\texttt{Maj} $>$ \texttt{Maj\_PSF}  & \numprint{7211}  & 78\% \\
\hline
\end{tabular}}
\tablefoot{$N_\mathrm{g}$ represents the number of sources remaining after each cut, along with the corresponding percentage of the initial sample. "Resolved" indicates sources with non-zero deconvolved major and minor axes. All sources have peak flux S/N larger than 5, which is the S/N threshold used when creating the raw catalogue (see Sect. 5 in \citealp{deJong2024} for more information).}
\end{table}

In both the LoTSS and ILT samples, we do not explicitly separate AGNs and SFGs because our selection criteria naturally result in an SFG-dominated population. An analysis of the LoTSS Deep Fields Data Release 1 \citep{Best2023} found that $\sim$80\% of sources are SFGs. Given the improved sensitivity in LoTSS Deep Fields Data Release 2 and ILT observations compared to LoTSS Deep Fields Data Release 1, we expect an even higher SFG fraction in our samples. Besides, by selecting only resolved sources with simple morphologies, we have excluded most AGNs, which often exhibit complex structures poorly fit by Gaussian profiles. A SFG-dominated sample with simple source shapes is sufficient for the purposes of this study, although residual AGN contamination may still introduce a small model bias in weak lensing analysis.

\subsection{HSC-SSP survey data}
\label{subsec:hsc}

HSC-SSP is an ongoing multi-band imaging survey using the Hyper Suprime-Cam instrument on the Subaru 8.2m telescope \citep{Aihara2018a}. The survey consists of three layers going to different depths: Wide, Deep, and UltraDeep, of which the Wide layer is specifically designed for the weak lensing analysis, the primary science driver of the HSC-SSP survey. The ELAIS-N1 field is one of the HSC Deep layer fields. It does not lie within the Wide layer footprint, and was therefore excluded in their published cosmic shear analyses \citep[e.g.][]{Hikage2019, Hamana2020, Dalal2023}. However, we will show later in Sect.~\ref{sec:hsc_2pcf} that, benefiting from the excellent image quality, a statistically significant cosmic shear signal can be detected in this field with HSC data.

In our work, we employed the HSC-SSP S16A data release which was made public as part of HSC-SSP second data release \citep[see][]{Aihara2019}. This release provides photometric data in five broad bands ($grizy$) along with two other narrow bands for the ELAIS-N1 field. The $i$-band images, with the given priority in the observing strategy, have the highest image quality and as such were used for the measurement of galaxy shapes. The HSC image and catalogue data we used in this paper are the outputs from the HSC pipeline \citep{Bosch2018}.

We imposed the same galaxy selection criteria on the S16A ELAIS-N1 catalogue as in the HSC first-year (Y1) shear catalogue paper \citep{Mandelbaum2018Jan}. This ensures the selection of a galaxy sample with nearly identical data quality as the Y1 shear catalogue\footnote{HSC Y1 shear catalogue is based on the Wide S16A dataset and is available for direct download from the HSC postgreSQL database. Instructions can be found here: \url{https://hsc-release.mtk.nao.ac.jp/doc/index.php/s16a-shape-catalog-pdr2/}.}, essential to apply the HSC open-source shear calibration software\footnote{\url{https://github.com/PrincetonUniversity/hsc-y1-shear-calib/}}. We define the weak lensing full depth and full colour (WLFDFC) area for the HSC ELAIS-N1 field, requiring that the number of visits \texttt{countInputs} for each $grizy$ band is larger than 3, 3, 4, 4, 4 visits respectively. To mask out the sources near the bright objects, we set \texttt{iflags\_pixel\_bright\_object\_any=False}. In addition, the $i$-band CModel\footnote{Composite model photometry, the primary photometry algorithm used in the HSC pipeline.} magnitude of the weak lensing sources should be lower than 24.5 AB mag after corrected for extinction. Refer to Sect.~5.1 or Table 4 in \cite{Mandelbaum2018Jan} for a full description of all the galaxy cuts.

\begin{figure}[htbp]
    \includegraphics{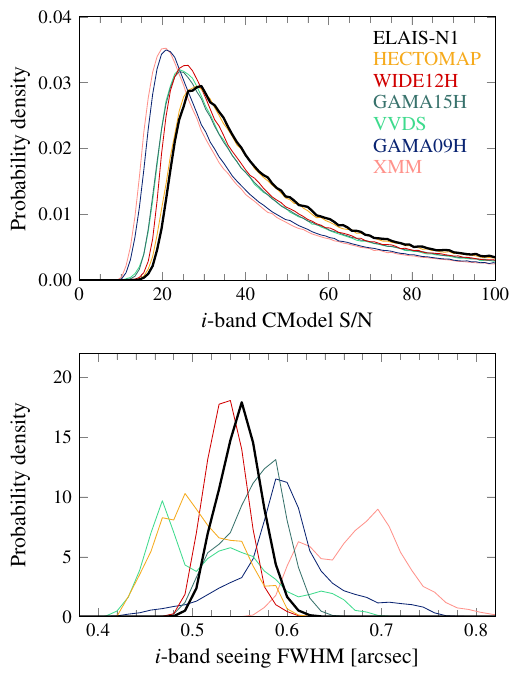}
    \caption{Unweighted distributions of the $i$-band CModel S/N and seeing FWHM for the galaxies in the HSC ELAIS-N1 weak lensing sample and in the six disjointed fields of the HSC Y1 shear catalogue (XMM, GAMA15H, GAMA09H, HECTOMAP, VVDS, and WIDE12H). The seeing FWHM is calculated assuming a Gaussian PSF.}
    \label{fig:hsc}
\end{figure}

The final HSC ELAIS-N1 weak lensing sample contains \numprint{456593} sources and covers a total area of 6.4 deg$^2$, reaching a 5$\sigma$ point source depth of 26.6 AB mag and mean seeing of $\sim$0.54 arcsec in the $i$-band. We estimate the 5$\sigma$ limiting magnitude for point source by averaging the PSF magnitudes of stars with a S/N between 4.9 and 5.1 in the ELAIS-N1 field after applying area cuts. This estimate is considered reasonable in this field for the S16A data release, which includes 90 nights of observations, as it lies in between the $i$-band depths of the same field in the HSC-SSP first and second data releases. Their depths are 26.5 and 26.6 mag, respectively, with observing time of 61.5 and 174 nights \citep{Aihara2018b, Aihara2019}. In Fig.~\ref{fig:hsc}, we compare the two key sample characteristics relevant to shear calibration: seeing conditions and the S/N distributions, between the ELAIS-N1 weak lensing sample and the Y1 shear catalogue. The S/N distribution of the ELAIS-N1 sample (black line) resembles that of the HECTOMAP field in the Y1 shear catalogue (faint yellow line), and the seeing condition is among those in the Y1 shear catalogue. The differences are very minor, suggesting that the aforementioned shear calibration code used for the Y1 shear catalogue remains applicable to the ELAIS-N1 sample and would not cause a serious problem in the subsequent analysis. Note that the S/Ns of the galaxies in ELAIS-N1 are in general higher compared to those in the Y1 shear catalogue, as shown in the top panel of Fig.~\ref{fig:hsc}. The former goes approximately 0.6 mag deeper on average than the latter catalogue.

The galaxy shapes in the ELAIS-N1 weak lensing sample are obtained using re-Gaussianization PSF correction method based on the measurements of moments \citep{Hirata2003}. Following Appendix 3 of \cite{Mandelbaum2018Jan}, the shear estimate, $\hat{\bm{\gamma}}=(\hat{\gamma}_1, \hat{\gamma}_2)$, for each galaxy is defined as
\begin{equation}\label{eq:shear}
    \hat{\bm{\gamma}}=\frac{1}{1+\langle m\rangle}\left(\frac{\bm{e}}{2\mathcal{R}}-\bm{c}\right),
\end{equation}
where $\bm{e}=(e_1, e_2)$ is the measured ellipticity, $\bm{c}=(c_1, c_2)$ is the additive bias, $\langle m\rangle$ denotes the weighted average of the multiplicative bias, with the weight given by
\begin{equation}
    w=\frac{1}{\sigma_e^2+e_\mathrm{rms}^2},
\end{equation}
where $\sigma_e$ denotes the shape measurement error, and $e_\mathrm{rms}$ is the intrinsic shape dispersion. $\mathcal{R}$ represents the shear response of the measured ellipticity to a small shear distortion \citep{Kaiser1995, Bernstein2002},
\begin{equation}
    \mathcal{R}=1-\langle e_\mathrm{rms}^2\rangle.
\end{equation}
The quantities in Eq.~(\ref{eq:shear}) are either available in the HSC public database or can be calculated from the additional data columns generated by the shear calibration software, where the shape noise and calibration factors are derived using the HSC weak lensing simulations described in \cite{Mandelbaum2018Dec}.

We assign a redshift to each galaxy in the ELAIS-N1 weak lensing sample with the best point estimate of the photometric redshift (photo-$z$) derived from a neural network code, \textsc{Ephor AB} \citep[see][]{Tanaka2018}. For the estimate of the redshift probability distribution function (PDF) of the sample, we employ a mathematically convenient method provided in \cite{Tanaka2018}, Gaussian Kernel Density Estimator (KDE),
\begin{equation}
    n^\mathrm{MC}(z)=\frac{1}{\sqrt{2\pi}Nh}\sum_i^N \exp\left[-\frac{(z-z_{\mathrm{MC},i})^2}{2h^2}\right],
\end{equation}
where $h=0.05$ is the kernel width for the \textsc{Ephor AB} code, $N$ denotes the total number of sample galaxies, and $z_\mathrm{MC}$ is a Monte Carlo draw from the redshift probability distribution for each galaxy, provided in the HSC database with column name \texttt{photoz\_mc}. The PDF inferred by the Gaussian KDE is in good agreement with both the stacked photo-$z$ PDF, a sum of all the photo-$z$ probability distributions of individual galaxies, and the PDF obtained by reweighting the reference redshift sample, as has been demonstrated in \cite{Tanaka2018}.

\subsection{Radio-optical cross-matching}
\label{subsec:cross}

In the ELAIS-N1 field, 99\% of radio detections in the LoTSS Deep Fields DR2 survey have identified optical or near-infrared counterparts (Bisigello et al., in prep.). This matching fraction should also applies to the ILT survey, as its catalogue has already been cross-matched with LoTSS data \citep{deJong2024}.

In this study, however, we performed a cross-matching between the radio samples and the HSC weak lensing catalogue, which only includes sources with sufficient S/N and sizes for weak lensing measurements. Since the weak lensing catalogue is a subset of all HSC optical detections, many radio sources remain unmatched. For both the full catalogue from LoTSS Deep Field survey and the selected samples from ILT data, we used a matching radius of $1^{\prime\prime}$. Out of \numprint{57594} radio sources in the overlapping region of the HSC and LoTSS Deep Fields, \numprint{21845} were successfully matched. For ILT data, \numprint{2595} sources were matched out of \numprint{6259} sources in the overlapped area.

The matching fractions for LoTSS and ILT data with matched HSC sources are 38\% and 41\% respectively, which are slightly lower than the matching fraction reported by SuperCLASS survey\footnote{In SuperCLASS survey analysis, an matching fraction of $\sim$43\% ($\sim$62\%) was obtained by cross-matching the \textit{e}-MERLIN (JVLA) radio observations with the HSC-SSP DR1 weak lensing catalogue. The noise levels for the \textit{e}-MERLIN and JVLA observations are 7 $\mathrm{\mu Jy\ beam^{-1}}$ at $L$-band (1.4 GHz). When converted to the equivalent noise level at 150 MHz, using a spectral index of 0.7, this corresponds to approximately 33 $\mathrm{\mu Jy\ beam^{-1}}$.} \citep{Battye2020}. The radio observations in the SuperCLASS survey are less sensitive than those from LoTSS or the ILT. And since the HSC DR1 data used in their cross-matching analysis has a depth comparable to the HSC data employed in our study, the lower matching fraction we report is consistent with expectations.

\section{Two-point shear correlation functions in the ELAIS-N1 field with HSC data}
\label{sec:hsc_2pcf}

\begin{figure}
    \includegraphics{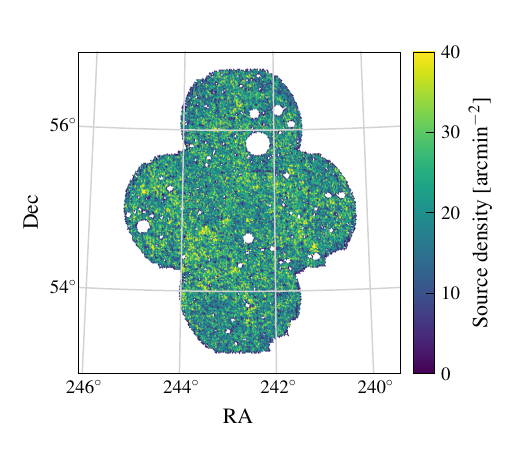}
    \caption{Unweighted source density distribution of the weak lensing sources in the ELAIS-N1 field. The unweighted mean source density for all the sources that pass the galaxy cuts in Sect.~\ref{subsec:hsc} is 19.7 $\mathrm{arcmin}^{-2}$. This plot was generated with a \textsc{HEALPix} pixelisation parameter of $N_\mathrm{side}=4096$.}
    \label{fig:source_density_distribution}
\end{figure}

\begin{figure}[htbp]
    \includegraphics{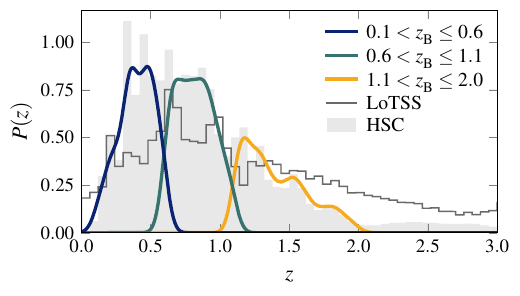}
    \caption{Photometric redshift distribution. The solid lines represent the weighted redshift PDFs for the three tomographic redshift bins: $0.1<z_\mathrm{B}\leq0.6$, $0.6<z_\mathrm{B}\leq1.1$, and $1.1<z_\mathrm{B}\leq2.0$ from the HSC weak lensing catalogue. The light-grey histogram shows the weighted distribution of the best point estimates of the photometric redshift for all the HSC sources in the ELAIS-N1 weak lensing sample. The dark-grey histogram shows the redshift distribution of the LoTSS sources using photometric redshift from matched Euclid Deep Field North data.}
    \label{fig:Pz_ELAIS-N1}
\end{figure}

We present the shear 2pCF analysis results with the HSC optical data alone in this section. After imposing the galaxy cuts in Sect.~\ref{subsec:hsc}, the remaining galaxies have a density distribution shown in Fig.~\ref{fig:source_density_distribution}. For the measurement of the shear 2pCFs, we use only the galaxies whose best estimate redshift $z_\mathrm{B}$ is within the range from 0.1 to 2.0, as the low redshift sources contain little lensing signal and the sources with $z_\mathrm{B}>2$ are detected with relatively low S/Ns. These galaxies are then divided into 3 tomographic redshift bins with bin edges [0.1, 0.6, 1.1, 2.0]. Figure~\ref{fig:Pz_ELAIS-N1} shows the redshift distributions of the galaxies in the three tomographic bins. The median redshift for these tomographic bins are 0.39, 0.81, and 1.35 respectively. The overall usable galaxy number density ($0.1<z_\mathrm{B}\leq2.0$) is 19.2 arcmin$^{-2}$. More properties regarding the individual tomographic sample are listed in Table~\ref{table:hsc_sample_properties}.

\begin{table}[htbp]
\caption{Sample properties for the three tomographic bins.}
\centering
{\renewcommand{\arraystretch}{1.3}
\label{table:hsc_sample_properties}
\begin{tabular}{c c c c c}
\hline\hline
$z$ range & $z_\mathrm{med}$ & $N_\mathrm{g}$ & $n_\mathrm{g,eff}$ & $\langle e_\mathrm{rms}^2\rangle^{1/2}$  \\
 & & & [arcmin$^{-2}$] & \\
\hline
$0.1<z\leq0.6$ & 0.392 & \numprint{148284} &  6.7 & 0.393 \\
$0.6<z\leq1.1$ & 0.813 & \numprint{163117} &  7.3 & 0.395 \\
$1.1<z\leq2.0$ & 1.347 & \numprint{110208} &  5.2 & 0.407 \\
\hline
$1.1<z\leq2.0$ & 0.742 & \numprint{421609} & 18.3 & 0.397 \\
\hline
\end{tabular}}
\tablefoot{$N_\mathrm{g}$ is the total number of sources. $n_\mathrm{g,eff}$ is the effective galaxy number density using the definition in \cite{Heymans2012}. $z_\mathrm{med}$ and $\langle e_\mathrm{rms}^2 \rangle^{1/2}$ are the weighted median redshift and the weighted mean intrinsic dispersion, respectively.}
\end{table}

The shear 2pCFs can be estimated from
\begin{equation}
    \hat{\xi}_\pm(\theta) = \frac{\sum_{ij} w_i w_j\ [\hat{\gamma}_{t}(\boldsymbol{\theta}_i)\hat{\gamma}_{t}(\boldsymbol{\theta}_j) \pm \hat{\gamma}_{\times}(\boldsymbol{\theta}_i)\hat{\gamma}_{\times}(\boldsymbol{\theta}_j)] }{\sum_{ij} w_i w_j},
\end{equation}
where the summation runs over the galaxy pairs within an angular separation bin. The 2pCFs are measured in a angular range from 1$^\prime$ to 80$^\prime$, using the public software \textsc{TreeCorr} \citep{Jarvis2004}. We present only the measured $\hat{\xi}_+$ in this paper, because the detection of $\hat{\xi}_-$ is not significant. In Appendix~\ref{appendix:psf_leakage}, we show that the systematic from PSF leakage and resulting PSF contribution to the shear $\xi_+$ is less than $10^{-6}$, which is well below the lensing signal. As the ELAIS-N1 field is relatively small, we do not expect to achieve a very precise measurement of the shear 2pCF signal, and thus corrections for residual PSF effects are not necessary in our study. Across the full redshift range $0.1 < z \leq 2.0$, $\xi_+$ is detected at a significance level of $\sim$9$\sigma$.

\begin{figure*}[htbp]
    \centering
    \includegraphics{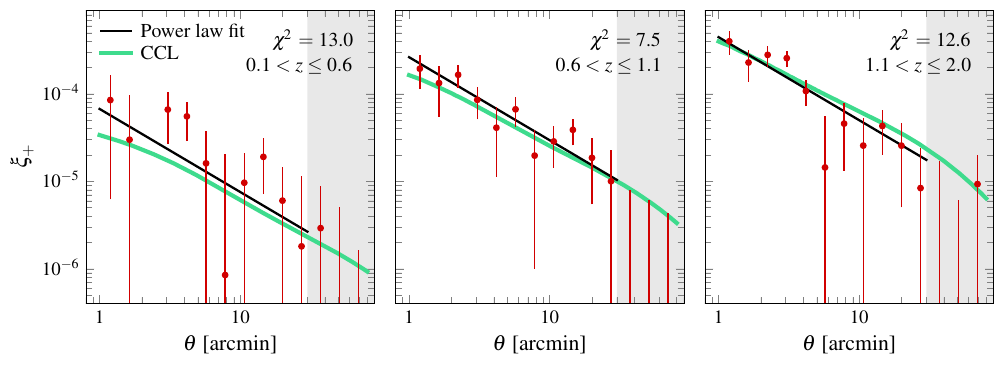}
    \caption{Tomographic cosmic shear 2pCFs (red) with the power law fitting lines (black) using a slope of $\gamma=0.95$. The error bars are calculated from bootstrap resampling using \textsc{TreeCorr}. Measurements in the grey region ($\theta>30^\prime$) are excluded from the fitting due to the increasing sample variance and higher contamination from PSF leakage. Green lines are theory 2pCFs from \textsc{PyCCL}. Best-fit $\chi^2$ values for the power-law model are indicated in the top-right corner of each panel. The 2pCF amplitudes at the angular separation $\theta_\mathrm{p}=3.2^\prime$ are $0.20\pm0.15$, $0.87\pm0.10$, and $1.53\pm0.17$ for respectively $0.1<z<\leq0.6$, $0.6<z\leq1.1$, and $1.1<z\leq2.0$. There is a clear trend that the lensing signal becomes stronger at higher redshifts.}
    \label{fig:hsc_tpcf_xip}
\end{figure*}

To determine the amplitude of the measured lensing signal, we fit the 2pCFs with a power law function,
\begin{equation}
    \xi_+(\theta)=A\left(\frac{\theta}{\theta_\mathrm{p}}\right)^{-\gamma},
\end{equation}
where $\gamma$ is the power law index, $\theta_\mathrm{p}$ is the pivot angular scale at which we evaluate the amplitude $A$ of the shear correlation signal. $\theta_\mathrm{p}$ is determined such that the degeneracy between parameters $\gamma$ and $A$ is minimised\footnote{This is done by tuning $\theta_\mathrm{p}$ and minimising the covariance between $\gamma$ and $A$.}. Using the galaxy samples from the two higher redshift bins, we find $\theta_\mathrm{p}=3.2^\prime$ and the best-fit $\gamma=0.95$. We then apply the power law fitting to the three tomographic bins while keeping $\theta_\mathrm{p}$ and $\gamma$ fixed. The amplitudes of $\xi_+$ for the three bins are presented in the second column of Table~\ref{table:hsc_2pCF_amp}. In the redshift bins $0.6<z\leq1.1$ and $1.1<z\leq2.0$, the $\xi_+$ amplitudes are measured with high statistical significance. Fig.~\ref{fig:hsc_tpcf_xip} shows the measured 2pCFs together with the best-fit power law functions for three redshift ranges. We note that there is a clear redshift dependence of the measured correlations, and the 2pCFs are consistent with the prediction by \textsc{PyCCL}\footnote{\url{https://github.com/LSSTDESC/CCL}} \citep{Chisari2019}.

We also conducted a 2pCF analysis using the LoTSS-matched HSC sources. The cross-matched sample comprises \numprint{21845} sources, with a galaxy density of $\sim$1 arcmin$^{-2}$ and a median redshift of $z_\mathrm{med}=0.86$ (see in Table~\ref{table:summary statistics} for a comparison of summary statistics). To mitigate systematics from sample variance at large separations, our measurements are restricted to the angular separation range $1^\prime < \theta < 20^\prime$. The results are shown in the third column of Table~\ref{table:hsc_2pCF_amp}. The tomographic $\xi_+$ measurements are dominated by noise and hence unreliable. The non-tomographic $\xi_+$ amplitude over redshift range $0.1<z\leq2.0$ reaches a significance level of $\sim$2$\sigma$. Notably, although the LoTSS-matched sample exhibits a higher median redshift than the HSC data, the cross-matching reduces the redshift advantage provided by the radio survey. According to host identification analysis of LoTSS sources using Euclid Deep Field North data by Bisigello et al. (in prep.), LoTSS spans a broader range of redshifts than HSC lensing samples, as shown in Fig.~\ref{fig:Pz_ELAIS-N1}. Matching with a deeper optical lensing survey would result in a higher median redshift and an increased source count, both of which would increase the S/N. This analysis demonstrates the potential of radio surveys as deep as LoTSS for cosmic shear studies, provided the resolution were increased.

\begin{table}[htbp]
\caption{Amplitudes of $\xi_+$ evaluated at $\theta_\mathrm{p}=3.2^\prime$ using HSC deep layer data in ELAIS-N1 field.}
\centering
{\renewcommand{\arraystretch}{1.3}
\label{table:hsc_2pCF_amp}
\begin{tabular}{c c c}
\hline\hline
\multirow{2}{*}{$z$ range} & \multicolumn{2}{c}{$A\times10^4$} \\ \cline{2-3} 
                           & full           & LoTSS-matched \\ \hline
$0.1<z\leq0.6$ & $0.20\pm0.15$ & $1.86\pm1.96$     \\
$0.6<z\leq1.1$ & $0.87\pm0.10$ & $3.33\pm1.25$     \\
$1.1<z\leq2.0$ & $1.53\pm0.17$ & $1.95\pm1.39$     \\ \hline
$0.1<z\leq2.0$ & $0.56\pm0.03$ & $0.69\pm0.33$     \\ \hline
\end{tabular}}
\tablefoot{The 2pCFs are computed within a separation range of $1^\prime < \theta < 30^\prime$ for the full HSC sample and $1^\prime < \theta < 20^\prime$ for the LoTSS-matched HSC sample.}
\end{table}

\section{Comparison of radio and optical samples}
\label{sec:compare}

\subsection{Summary statistics}
\label{subsec:sample statistics}

Table~\ref{table:summary statistics} summarises the properties of the samples that we have selected in Sect.~\ref{sec:data}. We note that:
\begin{itemize}
    \item Though the source density in LoTSS Deep Fields DR2, 1.75 arcmin$^{-2}$, is relatively high for a radio survey, the majority of sources are unresolved due to the low resolution, rendering them unusable for weak lensing measurements. In the case of ILT sub-arcsecond sources, a significant number of resolved sources are maintained (see Table~\ref{table:cuts on lofar high res}); however, the number density, 0.3 arcmin$^{-2}$, remains insufficient for reliable cosmic shear detection. The detection of shear signal typically requires at least a few galaxies per square arcmin.
    \item In the cross-matched HSC × ILT sub-arcsecond sample, the radio shape measurement noise ($\sigma_e=0.227$) is approximately 3 times higher than the HSC shape measurement noise ($\sigma_e=0.083$), but this is still smaller than the intrinsic noise and therefore the shapes are not measurement noise dominated.
    \item The median redshifts of both the HSC × LoTSS DR2 and HSC × ILT sub-arcsecond samples are higher than that of the HSC, highlighting a potential advantage of radio weak lensing over optical lensing.
\end{itemize}
The current low source number density and high measurement noise do not allow us to conduct weak lensing shear analysis with LoTSS or ILT data. We place our hope in future deeper surveys, where more sources will be detected and shape measurement noise will be significantly reduced.

\begin{table*}[htbp]
\caption{Summary statistics of radio and optical samples in ELAIS-N1 field.}
\centering
{\renewcommand{\arraystretch}{1.3}
\label{table:summary statistics}
\begin{tabular}{l r r r c c c c c}
\hline\hline
Sample & $N_\mathrm{g}$ & Area [deg$^2$] & $n_\mathrm{g}$ [arcmin$^{-2}$] & $z_\mathrm{med}$  & $z_\mathrm{mean}$  & $\langle e_\mathrm{rms}^2\rangle^{1/2}$ & $\langle\sigma_e^2 \rangle^{1/2}$ & $\langle e_\mathrm{rms}^2+\sigma_e^2 \rangle^{1/2}$ \\
\hline
HSC                             & \numprint{456593} &  6.4  & 19.71 & 0.81 & 0.97 & \textbf{0.402} & \textbf{0.130} & \textbf{0.423} \\
LoTSS DR2 full                  & \numprint{154952} &  24.5 & 1.76  & -    & -    & -     & -     & -    \\
ILT sub-arcsecond               & \numprint{7211}   &  6.7  & 0.30  & -    & -    & 0.365 & 0.236 & 0.435 \\
\hline
HSC × LoTSS DR2 full            & \numprint{21845} &  6.4  & 0.95  & 0.84 & 0.86 & \textbf{0.408} & \textbf{0.075} & \textbf{0.415} \\
HSC × ILT sub-arcsecond         & \numprint{2595}   &  5.2  & 0.14  & 0.91 & 0.92 & \textbf{0.399} & \textbf{0.083} & \textbf{0.407} \\
\ \ \ \ \large{\rotatebox[origin=c]{180}{$\Lsh$}}\normalsize{ Radio shape} 
                                & -                 &  -    & -     & -    & -    & 0.362 & 0.227 & 0.427 \\
\hline
\end{tabular}}
\tablefoot{$N_\mathrm{g}$ and $n_\mathrm{g}$ are the total number of sources and mean number density. All median and mean values are unweighted. Values in bold indicate the optical shape measurements from the HSC. $\langle e_\mathrm{rms}^2\rangle^{1/2}$ and $\langle\sigma_e^2 \rangle^{1/2}$ represent the intrinsic shape dispersion and measurement uncertainty, respectively. $\langle e_\mathrm{rms}^2 + \sigma_e^2 \rangle^{1/2}$ denotes the total ellipticity dispersion. For ILT data, the total ellipticity dispersion $\langle e_\mathrm{rms}^2 + \sigma_e^2 \rangle^{1/2}$ is measured directly, the measurement uncertainty $\langle\sigma_e^2 \rangle^{1/2}$ is estimated through Eq.~(\ref{eq:ellipticity}), and the intrinsic shape dispersion $\langle e_\mathrm{rms}^2\rangle^{1/2}$ is derived from these quantities. Details on the LoTSS DR2, ILT sub-arcsecond, and HSC samples are provided in Sect.~\ref{subsec:lofar}, Sect.~\ref{subsec:ILT}, and Sect.~\ref{subsec:hsc}.}
\end{table*}

\subsection{Comparison of radio and optical shapes}
\label{subsec:shape_comparison}

\begin{figure}
    \centering
    \includegraphics{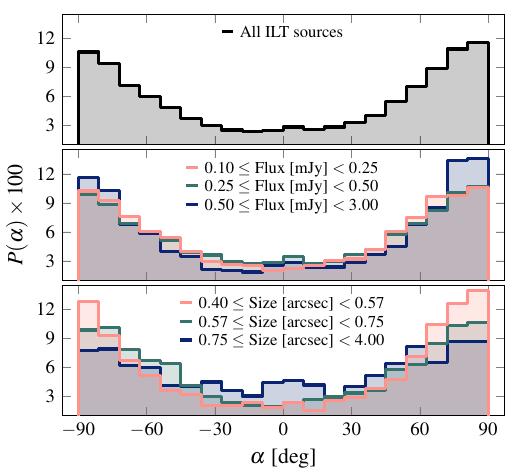}
    \caption{Position angle distributions for all ILT sources (top), flux-binned sub-samples (middle), and size-binned sub-samples (bottom). While PSF contamination is present across all samples, its strength shows a clear dependence on source size -- decreasing systematically with increasing angular size.}
    \label{fig:pa_hist}
\end{figure}

We explore the radio-optical shape correlation using the matched ILT samples in Sect.~\ref{subsec:cross}. Since the radio emission of SFGs traces the star-forming regions, a positive radio-optical shape correlation is expected if the measured radio shapes in LOFAR observations are not biased by systematics.

\begin{figure}
    \centering
    \includegraphics{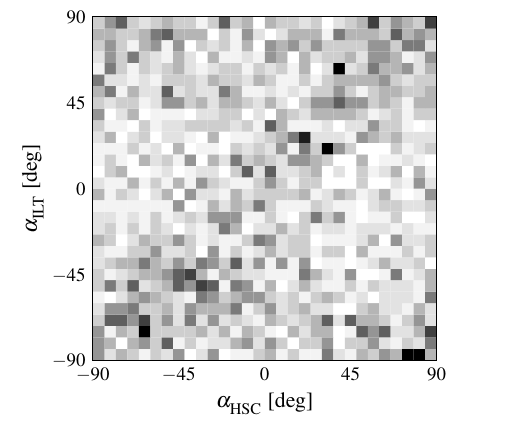}
    \caption{2D histogram of the radio and optical position angles of the cross-matched catalogue between ILT sub-arcsecond and HSC.}
    \label{fig:pa_2Dhist}
\end{figure}

\begin{figure}
    \centering
    \includegraphics{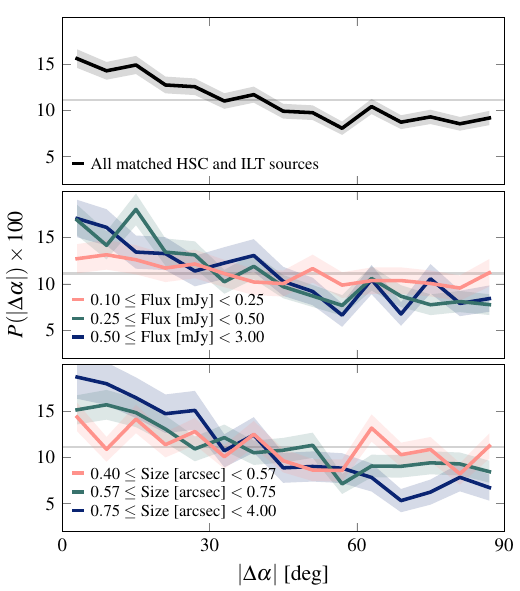}
    \caption{Distributions of the absolute difference in position angles between matched ILT and HSC sources. The top panel shows the overall distribution for all matched sources. The lower panels show distributions for sub-samples divided by radio flux density (middle) and by radio size (bottom). The shaded error band reflects Poisson uncertainties. The solid grey line represents the scenario where there is no correlation between the radio and optical position angles. For all matched sources, the KS statistic relative to a uniform distribution is $D=0.10$. For the 3 increasing flux bins, the KS statistics are $D=0.06, 0.12, 0.12$. For the 3 increasing size bins, the KS statistics are $D=0.06, 0.10, 0.18$. In all cases, the $p$-values are smaller than 0.005.}
    \label{fig:delta_pa}
\end{figure}

Both HSC and LOFAR observed shapes (or shears) suffer from various systematics, with the PSF systematic being the most significant. For optical weak lensing surveys like HSC-SSP, the methodology for shear calibration is relatively mature, and the PSF residual left in the corrected shears is usually negligible compared to the cosmic shear signal. Nonetheless, for radio observations, even after deconvolution, significant PSF contamination remains evident. For instance, in ILT data, we find average ellipticity values of $\langle e_1 \rangle = -0.053$ and $\langle e_2 \rangle = 0.09$, which match the PSF ellipse shape (see Appendix~\ref{appendix:lofar_psf} for ILT PSF image). Further evidence of PSF contamination is seen in the position angle distribution (Fig.~\ref{fig:pa_hist}). In this paper, we do not correct the ILT measured shapes. Because our current goal is not to conduct radio weak lensing but to explore its capabilities with LOFAR data, precise calibration of radio shear would be premature.

To examine the correlation between radio and optical shapes, we utilise only the radio and optical orientations (position angles), as this approach is less sensitive to the intrinsic galaxy shapes. In Fig.~\ref{fig:pa_2Dhist}, we present the histogram of radio and optical position angles for the matched HSC × ILT sub-arcsecond sources. Evidence of positive correlation can be clearly observed along the diagonal line, as well as in the $(-90^\circ, 90^\circ)$ and $(90^\circ, -90^\circ)$ corners. This clear correlation in the HSC × ILT sub-arcsecond sample is also visible in the top panel of Fig.~\ref{fig:delta_pa}, where the $\Delta\alpha$ exhibits a peak at lower values and gradually decreases towards higher values. A Kolmogorov-Smirnov (KS) test against uniform distribution yields a statistic $D=0.10$ with a $p$-value of $2\times10^{-21}$, showing clear disagreement with a uniform distribution. We further split the sample into radio flux and radio size sub-samples and show the $|\Delta\alpha|$ distributions in the lower two panels of Fig.~\ref{fig:delta_pa}. Both high-flux and large-size radio sources exhibit position angles that are more closely aligned with their optical counterparts, with size having a stronger influence than flux.

Next, we need to provide a quantitative measure of the position angle correlation strength. Previous studies have preferred to use Pearson correlation coefficient of position angle directly to compare radio and optical shapes \citep[e.g.][]{Patel2010, Tunbridge2016, Hillier2019}. However, a linear correlation measure like that could not fully capture the relation between spin-2 quantities like position angles, due to their periodic nature. To address this limitation, we calculate the Pearson correlation coefficient between $\cos(2 \alpha_\mathrm{HSC})$ and $\cos(2\alpha_\mathrm{ILT})$. Alternatively, the correlation can also be evaluated using $\sin(2 \alpha_\mathrm{HSC})$ and $\sin(2\alpha_\mathrm{ILT})$. When the position angle $\alpha$ follows a uniform distribution, the correlation coefficients $R_{\sin(2\alpha)}$ and $R_{\cos(2\alpha)}$ are expected to be identical. Notably, in cases of strong correlation, our definition of the correlation coefficient converges to the traditionally used Pearson correlation coefficient.

\begin{figure}
    \centering
    \includegraphics{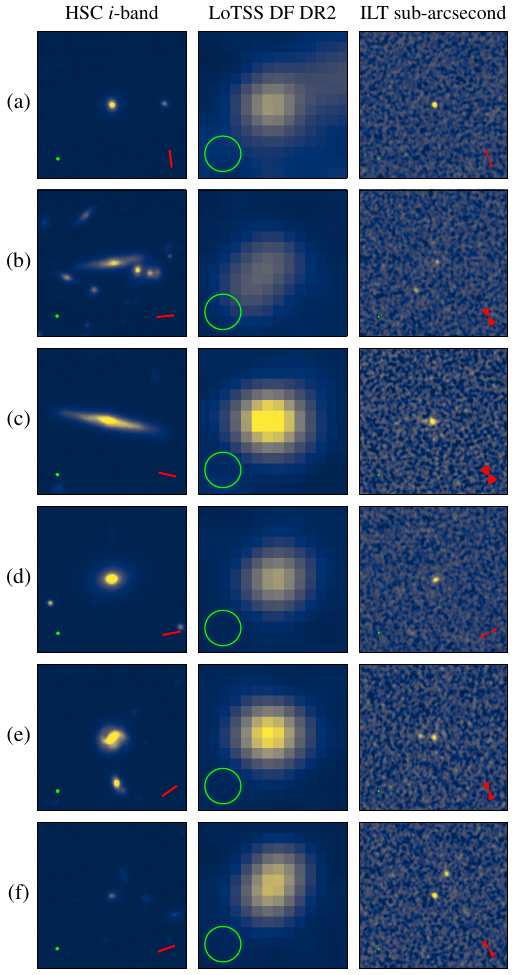}
    \caption{$21^{\prime\prime}\times21^{\prime\prime}$ cutout images for six sources in the cross-matched HSC, LoTSS Deep Field DR2 and ILT sub-arcsecond catalogues of ELAIS-N1. The LoTSS Deep Field DR2 images are restored with a circular Gaussian beam of FWHM $6^{\prime\prime}$, and the ILT sub-arcsecond images are restored with a elliptical Gaussian beam of FWHM $0.36^{\prime\prime}\times0.45^{\prime\prime}$. For better visualisation, the minimum and maximum pixel value are fixed to $[-1\sigma, 200\sigma]$, $[-3\sigma, 130\sigma]$, and $[-3\sigma, 10\sigma]$ for HSC, LoTSS and ILT data, respectively, where $\sigma$ represents the background RMS noise level for each image. The PSF sizes are shown in the bottom-left corner of each panel. In the HSC and ILT panels, the bottom-right corner shows the source position angles. For ILT sources, we also show the standard error of the position angle in a fan-shaped patch.}
    \label{fig:cutouts}
\end{figure}

We obtained $R_{\cos(2\alpha)}=0.15 \pm 0.02$, $R_{\sin(2\alpha)}=0.14 \pm 0.02$ for the matched sample, indicating a weak positive correlation. To illustrate the correlation, in Fig.~\ref{fig:cutouts}, we present images of 6 selected cross-matched sources between HSC, LoTSS and ILT. Although the ILT images lack some extended emission present in the HSC images due to limited depth, sources (a), (c), and (d) exhibit notable shape similarities between HSC and ILT. In contrast, the LoTSS sources in panel (b) and (f) appear to be two distinct sources blended together. This underscores the necessity of using ILT observations for shape measurements and highlights its potential for future radio weak lensing studies.

\section{Forecasts for International LOFAR Telescope}
\label{sec:forecast}

ILT offers significantly higher resolution compared to the Dutch LOFAR, enabling it to resolve sources that the Dutch array cannot. However, as we have shown above, the current ILT observations of ELAIS-N1 lack the necessary depth to be fully effective for weak lensing studies. In this section, we assess the potential of the ILT for weak lensing studies by considering two scenarios of deeper observations.

For both scenarios, we consider the same observing area as the ILT ELAIS-N1 data, which covers 6.7 deg$^2$, corresponding to a single ILT pointing. Based on our successful detection of cosmic shear in a similar sky region using HSC data (Sect.~\ref{sec:hsc_2pcf}), we anticipate that, with sufficient depth and an ILT resolution of 0.3$^{\prime\prime}$, detecting a cosmic shear signal should be achievable. We explore two integrated observing times: 128 hours and \numprint{3200} hours. A 32-hour ILT observation achieves a median RMS noise level of 17 $\mathrm{\mu Jy\ beam^{-1}}$ \citep{deJong2024}, and since the noise level is expected to scale with the inverse square root of the total observing time, we estimate the median noise levels for 128-hour and \numprint{3200}-hour observations to be approximately 8.5 and 1.7 $\mathrm{\mu Jy\ beam^{-1}}$, respectively.

We use the Tiered Radio Extragalactic Continuum Simulation (T-RECS; \citealp{Bonaldi2018}) to model the galaxy density and redshift distribution of radio populations. We employ the significance of total flux as the detection criterion and apply a total flux cut to the raw catalogue generated by T-RECS. Assuming a 10$\sigma$ detection threshold, the flux limits for the simulations are set at 85 $\mathrm{\mu Jy}$ and 17 $\mathrm{\mu Jy}$ for observation times of 128 hours and \numprint{3200} hours. From the simulated catalogues, we further select sources with angular sizes greater than the PSF size.\footnote{We set the PSF size to 0.4$^{\prime\prime}$, which corresponds to the geometric mean of the major and minor axes of the elliptical PSF from the 32-hour ILT ELAIS-N1 observation.} After applying these two cuts, the mock catalogues reach source densities of 2.0 arcmin$^{-2}$ for the 128-hour scenario and 6.5 arcmin$^{-2}$ for the \numprint{3200}-hour scenario.

\begin{figure}
    \centering
    \includegraphics{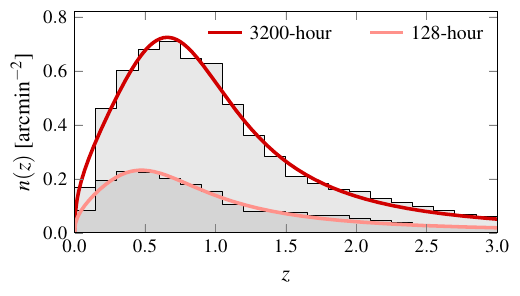}
    \caption{Redshift distributions of 128-hour (green) and \numprint{3200}-hour (red) mocks. Solid lines are best-fit redshift distribution model using the $n(z)$ parametrisation from \cite{Fu2008}. The redshift bin size is 0.15. The median redshifts for 128- and \numprint{3200}-hour simulations are 0.8 and 0.9.}
    \label{fig:mock_redshift}
\end{figure}

\begin{table}
\caption{Best-fit parameters of the model redshift distributions from Eq.~(\ref{eq:fu_nz}) for 128-hour and \numprint{3200}-hour mock catalogues.}
\centering
{\renewcommand{\arraystretch}{1.3}
\label{table:redshift_fit}
\begin{tabular}{c c c c c}
\hline\hline
Mock & $A$ [arcmin$^{-2}$] & $a$ & $b$ & $c$  \\
\hline
128-hour  & 0.457 & 0.441 & 3.923 & 0.637 \\
\numprint{3200}-hour & 0.093 & 0.410 & 3.014 & 0.352 \\
\hline
\end{tabular}}
\end{table}

\begin{figure}
    \centering
    \includegraphics{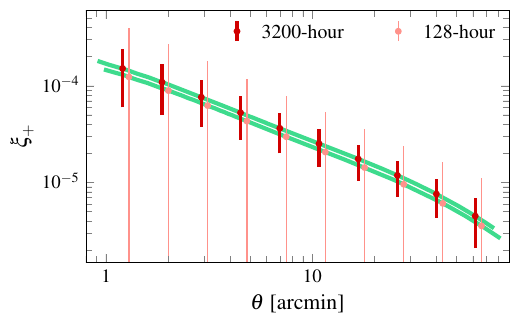}
    \caption{Weak lensing shear 2pCFs for forecasted 128- and \numprint{3200}-hour ILT observations. The error bars only include shot noise. Green lines indicate the theoretical 2pCFs predicted with \textsc{PyCCL} using the redshift distributions in Fig.~\ref{fig:mock_redshift}.}
    \label{fig:trecs_2pcf_xip}
\end{figure}

For redshift distributions, we smooth the raw $z$ distribution from mocks (shown in Fig.~\ref{fig:mock_redshift}) using $n_\mathrm{g}(z)$ model from \cite{Fu2008}:
\begin{equation}
\label{eq:fu_nz}
    n_\mathrm{g}(z)=A\frac{z^a+z^{ab}}{z^b+c},
\end{equation}
where $A$ is a normalisation parameter. The best-fit parameters are listed in Table~\ref{table:redshift_fit}. To make a simple prediction of shear 2pCF with the mock catalogues, we assume an intrinsic shape dispersion of $e_\mathrm{rms} = 0.3$ per component. Since the weak lensing 2pCF depends quadratically on galaxy ellipticities, the noise RMS for 2pCF signal is proportional to the square of the ellipticity dispersion. In this analysis, we only consider shot noise from shape dispersion. For each bin, the noise RMS is given by:
\begin{equation}
    \sigma_{\xi_+}(\theta) = \frac{\sqrt{2}\ e_\mathrm{rms}^2}{\sqrt{N_\mathrm{pair}(\theta)}}.
\end{equation}
Here, $N_\mathrm{pair}(\theta)$ is the number of galaxy pairs within a given redshift bin, derived directly from the mock catalogues. We use \textsc{PyCCL} to compute the theory shear 2pCFs. The weak lensing 2pCF forecasts for 128- and \numprint{3200}-hour ILT observations are shown in Fig.~\ref{fig:trecs_2pcf_xip}, with the 2pCF reaching significance levels of 1.8$\sigma$ and 6.8$\sigma$, respectively. Although these forecasts are optimistic, as they do not account for measurement noise or systematic errors, the results suggest that detecting weak lensing signals from ILT radio data may be feasible, particularly with \numprint{3200}-hour or deeper observations.

We should point out that in radio weak lensing, while increasing observation time helps reduce measurement noise, mitigating PSF systematics remains a major challenge. In radio interferometric imaging, the widely used \textsc{CLEAN} dirty image deconvolution algorithm \citep{Hogbom1974} performs suboptimally when applied to extended and diffuse emission, as it assumes the sources are composed of point sources. Therefore, this conventional deconvolution approach may introduce shear bias when extracting shape information from radio data. To address these limitations, new deconvolution algorithms have been developed that are better suited for reconstructing extended sources and thereby improving image quality. Yet, many of these advanced algorithms are computationally intensive in nature. Compressed sensing-based approaches \citep[e.g.][]{Carrillo2012, Carrillo2014, Dabbech2015}, which rely on sparsity assumptions, require global optimisation; Bayesian inference methods \citep[e.g.][]{Junklewitz2016} involve iterative probabilistic sampling; and neural networks \citep[e.g.][]{Connor2022} demand extensive training -- all significantly more computationally expensive than CLEAN's simpler, local reconstruction approach. Ongoing efforts to balance computational efficiency with the need for precise imaging are thus essential to fully unlock the potential of radio weak lensing as a tool for probing the universe's large-scale structure.

\section{Discussion}
\label{sec:discussion}

\subsection{Trade-off between resolution and sensitivity in radio surveys}
\label{subsec:trade-off}

Radio interferometers detect visibilities. By applying different weighting schemes to these visibilities, one can control the characteristics of PSF in the resulting image. In general, low-resolution imaging has better surface brightness sensitivity and is more effective at preserving extended emissions than high-resolution imaging, as it prioritises measurements from shorter baselines. \cite{deJong2024} offer ILT ELAIS-N1 catalogues at three resolutions: 0.3$^{\prime\prime}$, 0.6$^{\prime\prime}$, and 1.2$^{\prime\prime}$. In their analysis of integrated flux distribution across these 3 catalogues, they found that the 0.6$^{\prime\prime}$ and 1.2$^{\prime\prime}$ catalogues contain more detections than the 0.3$^{\prime\prime}$ catalogue at flux densities above $\sim$0.25 mJy and $\sim$0.55 mJy, respectively. Furthermore, around 20\% of the sources detected in the 1.2$^{\prime\prime}$ catalogue are missing from the 0.3$^{\prime\prime}$ catalogue. Since most of the detected sources are expected to be SFGs at the sensitivity level of ILT ELAIS-N1 data \citep{Best2023}, it is likely that a fraction of SFGs with sizes of less than an arcsecond account for the missing detections, as a resolution of 0.3$^{\prime\prime}$ tends resolve out these sources \citep{deJong2024}.

In this work, we have used the 0.3$^{\prime\prime}$ resolution ELAIS-N1 catalogue, as it provides the deepest map and yields the largest number of sources after selection, compared to the two lower resolutions. We have also verified that this resolution does not significantly resolve out extended emission when compared to 0.6\arcsec\ resolution measurements, retaining comparable shape characteristics. On average, the difference in source position angles measured at 0.3\arcsec\ and 0.6\arcsec\ resolution is 24$^\circ$  (see Appendix~\ref{appendix:ilt_weighting}). However, opting for the 0.3\arcsec resolution likely excluded a fraction of SFGs. For future radio weak lensing studies, it might be beneficial to optimize the trade-off between resolution and surface brightness sensitivity in order to maximize the number of detected sources available for weak lensing analysis.

\subsection{Radio shape measurements}

Throughout our work, we have used radio shape measurements from \textsc{PyBDSF}. Currently, \textsc{PyBDSF} is the most popular software for detecting and extracting source characteristics from radio images. It is particular suited for detecting radio sources as it can better handle the correlated noise inherent in radio images, whereas source extraction tools widely used in optical surveys like \textsc{SExtractor} \citep{Bertin1996}, assume uncorrelated noise. For characterising the shapes, \textsc{PyBDSF} models the source flux distribution as a group of 2D Gaussians, with the intention to capture the large-scale radio emission and complex morphologies typically associated with radio-loud AGNs, such as jets, lobes, and hotspots.

However, the \textsc{PyBDSF} shape measurement is not ideal for weak lensing purposes, as in deep radio observations, SFGs constitute the majority of the usable sources for weak lensing. SFGs often exhibit an exponential light profile, and fitting such a distribution with a 2D Gaussian can lead to systematic biases in ellipticity measurements. Moreover, the \textsc{PyBDSF} algorithm attempts to fit sources iteratively: it first fits a source with a single Gaussian, evaluates the residuals, and if the residual exceeds a predefined threshold, it adds additional Gaussians until an acceptable fit is achieved. As a result, high S/N SFGs are likely to be modelled with multiple Gaussians, further complicating the accurate determination of ellipticity. In radio weak lensing studies, where the extended emission are resolved to a certain degree, it is essential to separately measure their shapes to obtain more accurate measurements after \textsc{PyBDSF} identifies potential SFGs. For example, studies by \citet{Hillier2019}, \citet{Harrison2020}, and \citet{Tunbridge2016} used the \textsc{IM3SHAPE} image-plane shape measurement method \citep{Zuntz2013}, originally designed for optical data, in combination with a calibration for interferometric imaging artefacts, to derive radio shapes for selected sources. Alternatively, shapes can be measured in the visibility domain to avoid the highly non-linear imaging process \citep{Rivi2016, Rivi2018}. Nonetheless, these approaches have not yet been rigorously tested on real observations.

\subsection{Limitations of ILT on weak lensing}
\label{subsec:limitations}

In contrast to the SKA, the ILT is not designed as a survey telescope for large-scale cosmic structure studies. For instance, the proposed SKA Medium-Deep Band 2 Survey, which aims to study radio weak lensing, can cover approximately \numprint{5000} deg$^2$ in \numprint{10000} observation hours, achieving a usable source density of $\sim$3 sources $\mathrm{arcmin}^{-2}$ \citep{SquareKilometreArrayCosmologyScienceWorkingGroup2020}. The ILT, on the other hand, is optimized for deep, targeted observations, which limits its efficiency in achieving high source densities within short observation times. Weak lensing studies with the ILT can only be conducted in conjunction with ultra-deep surveys.

A second potential limitation for the ILT is DDEs of radio observations, with ionospheric distortions being the most severe one for LOFAR due to its very low operating frequency. Typically, direction-dependent calibration techniques are employed to address these issues. However, accurately obtaining calibrated visibilities requires a precise sky model, especially when accounting for ionospheric effects. Perfectly modelling the ionosphere is challenging, and the non-linear nature of the CLEAN deconvolution algorithm further complicates the reconstruction of source shapes. Combined, these factors make it uncertain how significantly ionospheric effect, radio calibration and imaging may influence the PSF systematics and thus affect weak lensing shape measurements.

\section{Conclusions}
\label{sec:conclusions}

In this paper, we present a weak lensing analysis alongside a radio-optical shape correlation analysis conducted in the ELAIS-N1 field. Our optical data in ELAIS-N1 are from HSC-SSP deep survey. We filtered the catalogue with exact same cuts as in the HSC weak lensing pipeline \citep{Mandelbaum2018Jan} and found a $\sim$$9\sigma$ detection of cosmic shear signal in the optical data via a 2pCF analysis. By dividing the lensing sources into three redshift bins, we also identified a clear signal indicative of its redshift dependence. This successful measurement raises the prospect of detecting similar signals using radio data on the same sources, even over relatively small sky areas.

We utilised two radio datasets from LOFAR in the ELAIS-N1 field: the LoTSS and ILT observations. The LoTSS survey provides the deepest low-frequency imaging of the field, with a source density reaching $\sim$2.7 arcmin$^{-2}$ in the central region. Using LoTSS-matched HSC samples, we measured the amplitude of shear correlation $\xi_+$ at a $\sim$2$\sigma$ significance level. While the LoTSS survey offers sufficient source density for weak lensing studies, its resolution is inadequate for precise shape measurements.

In contrast, the ILT survey delivers the highest-resolution radio imaging in ELAIS-N1 at 0.3\arcsec, but with a higher noise level compared to LoTSS. To access the potential of weak lensing analysis with ILT, we first examined the correlation between radio and optical shapes by cross-matching the ILT-detected sources with the HSC catalogue. The source position angle was used to evaluate the degree of radio-optical shape correlation. We measured a positive correlation of $R_{\cos(2\alpha)} = 0.15 \pm 0.02$. High-resolution ILT observations are therefore able to begin resolving SFGs.

Currently, the ILT survey depth is insufficient to achieve the source density required for a robust cosmic shear detection. The usable source density in the ILT ELAIS-N1 field is only 0.3 $\mathrm{arcmin}^{-2}$, over a factor of ten below what optical surveys can reach. Hence, we look forward to the future deeper ILT surveys and remain optimistic about the continued development of radio shear measurement techniques, which can take advantage of the high median redshift probed by radio surveys. We estimate that with \numprint{3200}-hour ILT observation over a single pointing ($\sim$6.7 deg$^2$), a $\sim$$6\sigma$ detection of the cosmic shear signal via 2pCF analysis could be achieved. At this depth, the source density would be adequate for weak lensing measurements. The primary challenge lies in controlling systematic errors. The ILT observations are affected by visibility interference from ionospheric effects, which disrupts the deterministic nature of the PSF. Additionally, the non-linear nature of the radio image deconvolution process complicates the reconstruction of radio shapes. We have shown in Sect.~\ref{subsec:shape_comparison} that the shapes from ILT observations are heavily contaminated by the PSF. Mitigating these systematics and accurate radio shape measurements will be critical for future weak lensing studies with radio data.

\begin{acknowledgements}

We thank Hironao Miyatake for helpful advice on using the HSC data products.

We thank George Miley, Jurjen de Jong, Timothy Shimwell for their helpful discussions.

We thank the anonymous referee for their helpful comments and suggestions.

JL is supported by the funding from China Scholarship Council (CSC) under the Agreement for Study Abroad for CSC Sponsored Chinese Citizens No 202206040036.

The Hyper Suprime-Cam (HSC) collaboration includes the astronomical communities of Japan and Taiwan, and Princeton University. The HSC instrumentation and software were developed by the National Astronomical Observatory of Japan (NAOJ), the Kavli Institute for the Physics and Mathematics of the Universe (Kavli IPMU), the University of Tokyo, the High Energy Accelerator Research Organization (KEK), the Academia Sinica Institute for Astronomy and Astrophysics in Taiwan (ASIAA), and Princeton University. Funding was contributed by the FIRST program from the Japanese Cabinet Office, the Ministry of Education, Culture, Sports, Science and Technology (MEXT), the Japan Society for the Promotion of Science (JSPS), Japan Science and Technology Agency (JST), the Toray Science Foundation, NAOJ, Kavli IPMU, KEK, ASIAA, and Princeton University.

This paper makes use of software developed for Vera C. Rubin Observatory. We thank the Rubin Observatory for making their code available as free software at \url{http://pipelines.lsst.io/}.

This paper is based on data collected at the Subaru Telescope and retrieved from the HSC data archive system, which is operated by the Subaru Telescope and Astronomy Data Center (ADC) at NAOJ. Data analysis was in part carried out with the cooperation of Center for Computational Astrophysics (CfCA), NAOJ. We are honoured and grateful for the opportunity of observing the Universe from Maunakea, which has the cultural, historical and natural significance in Hawaii.

LOFAR \citep{vanHaarlem2013} is the Low Frequency Array designed and constructed by ASTRON. It has observing, data processing, and data storage facilities in several countries, which are owned by various parties (each with their own funding sources), and that are collectively operated by the ILT foundation under a joint scientific policy. The ILT resources have benefited from the following recent major funding sources: CNRS-INSU, Observatoire de Paris and Université d'Orléans, France; BMBF, MIWF-NRW, MPG, Germany; Science Foundation Ireland (SFI), Department of Business, Enterprise and Innovation (DBEI), Ireland; NWO, The Netherlands; The Science and Technology Facilities Council, UK; Ministry of Science and Higher Education, Poland; The Istituto Nazionale di Astrofisica (INAF), Italy.

The data analysis was carried out with the use of \textsc{mocpy} \citep{Fernique2014}, \textsc{healpix} \citep{Gorski2005, Zonca2019}, \textsc{numpy} \citep{harris2020array}, \textsc{scipy} \citep{Virtanen2020SciPy}, \textsc{pandas} \citep{McKinney2010, pandas2020}, \textsc{astropy} \citep{astropy2013, astropy2018, astropy2022}, and \textsc{matplotlib} \citep{Hunter2007Matplotlib}.

\end{acknowledgements}

\bibliographystyle{bibtex/aa}
\bibliography{refs.bib}

\begin{thebibliography}{82}
\expandafter\ifx\csname natexlab\endcsname\relax\def\natexlab#1{#1}\fi

\bibitem[{{Abbott} {et~al.}(2018){Abbott}, {Abdalla}, {Alarcon}, {Aleksi{\'c}}, {Allam}, {Allen}, {Amara}, {Annis}, {Asorey}, {Avila}, \& et~al.}]{Abbott2018}
{Abbott}, T.~M.~C., {Abdalla}, F.~B., {Alarcon}, A., {et~al.} 2018, \prd, 98, 043526

\bibitem[{{Aihara} {et~al.}(2019){Aihara}, {AlSayyad}, {Ando}, {Armstrong}, {Bosch}, {Egami}, {Furusawa}, {Furusawa}, {Goulding}, {Harikane}, {Hikage}, {Ho}, {Hsieh}, {Huang}, {Ikeda}, {Imanishi}, {Ito}, {Iwata}, {Jaelani}, {Kakuma}, {Kawana}, {Kikuta}, {Kobayashi}, {Koike}, {Komiyama}, {Li}, {Liang}, {Lin}, {Luo}, {Lupton}, {Lust}, {MacArthur}, {Matsuoka}, {Mineo}, {Miyatake}, {Miyazaki}, {More}, {Murata}, {Namiki}, {Nishizawa}, {Oguri}, {Okabe}, {Okamoto}, {Okura}, {Ono}, {Onodera}, {Onoue}, {Osato}, {Ouchi}, {Shibuya}, {Strauss}, {Sugiyama}, {Suto}, {Takada}, {Takagi}, {Takata}, {Takita}, {Tanaka}, {Terai}, {Toba}, {Uchiyama}, {Utsumi}, {Wang}, {Wang}, \& {Yamada}}]{Aihara2019}
{Aihara}, H., {AlSayyad}, Y., {Ando}, M., {et~al.} 2019, \pasj, 71, 114

\bibitem[{{Aihara} {et~al.}(2018{\natexlab{a}}){Aihara}, {Arimoto}, {Armstrong}, {Arnouts}, {Bahcall}, {Bickerton}, {Bosch}, {Bundy}, {Capak}, {Chan}, {Chiba}, {Coupon}, {Egami}, {Enoki}, {Finet}, {Fujimori}, {Fujimoto}, {Furusawa}, {Furusawa}, {Goto}, {Goulding}, {Greco}, {Greene}, {Gunn}, {Hamana}, {Harikane}, {Hashimoto}, {Hattori}, {Hayashi}, {Hayashi}, {He{\l}miniak}, {Higuchi}, {Hikage}, {Ho}, {Hsieh}, {Huang}, {Huang}, {Ikeda}, {Imanishi}, {Inoue}, {Iwasawa}, {Iwata}, {Jaelani}, {Jian}, {Kamata}, {Karoji}, {Kashikawa}, {Katayama}, {Kawanomoto}, {Kayo}, {Koda}, {Koike}, {Kojima}, {Komiyama}, {Konno}, {Koshida}, {Koyama}, {Kusakabe}, {Leauthaud}, {Lee}, {Lin}, {Lin}, {Lupton}, {Mandelbaum}, {Matsuoka}, {Medezinski}, {Mineo}, {Miyama}, {Miyatake}, {Miyazaki}, {Momose}, {More}, {More}, {Moritani}, {Moriya}, {Morokuma}, {Mukae}, {Murata}, {Murayama}, {Nagao}, {Nakata}, {Niida}, {Niikura}, {Nishizawa}, {Obuchi}, {Oguri}, {Oishi}, {Okabe}, {Okamoto}, {Okura}, {Ono}, {Onodera}, {Onoue}, {Osato}, {Ouchi},
  {Price}, {Pyo}, {Sako}, {Sawicki}, {Shibuya}, {Shimasaku}, {Shimono}, {Shirasaki}, {Silverman}, {Simet}, {Speagle}, {Spergel}, {Strauss}, {Sugahara}, {Sugiyama}, {Suto}, {Suyu}, {Suzuki}, {Tait}, {Takada}, {Takata}, {Tamura}, {Tanaka}, {Tanaka}, {Tanaka}, {Tanaka}, {Terai}, {Terashima}, {Toba}, {Tominaga}, {Toshikawa}, {Turner}, {Uchida}, {Uchiyama}, {Umetsu}, {Uraguchi}, {Urata}, {Usuda}, {Utsumi}, {Wang}, {Wang}, {Wong}, {Yabe}, {Yamada}, {Yamanoi}, {Yasuda}, {Yeh}, {Yonehara}, \& {Yuma}}]{Aihara2018a}
{Aihara}, H., {Arimoto}, N., {Armstrong}, R., {et~al.} 2018{\natexlab{a}}, \pasj, 70, S4

\bibitem[{{Aihara} {et~al.}(2018{\natexlab{b}}){Aihara}, {Armstrong}, {Bickerton}, {Bosch}, {Coupon}, {Furusawa}, {Hayashi}, {Ikeda}, {Kamata}, {Karoji}, {Kawanomoto}, {Koike}, {Komiyama}, {Lang}, {Lupton}, {Mineo}, {Miyatake}, {Miyazaki}, {Morokuma}, {Obuchi}, {Oishi}, {Okura}, {Price}, {Takata}, {Tanaka}, {Tanaka}, {Tanaka}, {Uchida}, {Uraguchi}, {Utsumi}, {Wang}, {Yamada}, {Yamanoi}, {Yasuda}, {Arimoto}, {Chiba}, {Finet}, {Fujimori}, {Fujimoto}, {Furusawa}, {Goto}, {Goulding}, {Gunn}, {Harikane}, {Hattori}, {Hayashi}, {He{\l}miniak}, {Higuchi}, {Hikage}, {Ho}, {Hsieh}, {Huang}, {Huang}, {Imanishi}, {Iwata}, {Jaelani}, {Jian}, {Kashikawa}, {Katayama}, {Kojima}, {Konno}, {Koshida}, {Kusakabe}, {Leauthaud}, {Lee}, {Lin}, {Lin}, {Mandelbaum}, {Matsuoka}, {Medezinski}, {Miyama}, {Momose}, {More}, {More}, {Mukae}, {Murata}, {Murayama}, {Nagao}, {Nakata}, {Niida}, {Niikura}, {Nishizawa}, {Oguri}, {Okabe}, {Ono}, {Onodera}, {Onoue}, {Ouchi}, {Pyo}, {Shibuya}, {Shimasaku}, {Simet}, {Speagle}, {Spergel}, {Strauss},
  {Sugahara}, {Sugiyama}, {Suto}, {Suzuki}, {Tait}, {Takada}, {Terai}, {Toba}, {Turner}, {Uchiyama}, {Umetsu}, {Urata}, {Usuda}, {Yeh}, \& {Yuma}}]{Aihara2018b}
{Aihara}, H., {Armstrong}, R., {Bickerton}, S., {et~al.} 2018{\natexlab{b}}, \pasj, 70, S8

\bibitem[{{Amendola} {et~al.}(2018){Amendola}, {Appleby}, {Avgoustidis}, {Bacon}, {Baker}, {Baldi}, {Bartolo}, {Blanchard}, {Bonvin}, {Borgani}, {Branchini}, {Burrage}, {Camera}, {Carbone}, {Casarini}, {Cropper}, {de Rham}, {Dietrich}, {Di Porto}, {Durrer}, {Ealet}, {Ferreira}, {Finelli}, {Garc{\'\i}a-Bellido}, {Giannantonio}, {Guzzo}, {Heavens}, {Heisenberg}, {Heymans}, {Hoekstra}, {Hollenstein}, {Holmes}, {Hwang}, {Jahnke}, {Kitching}, {Koivisto}, {Kunz}, {La Vacca}, {Linder}, {March}, {Marra}, {Martins}, {Majerotto}, {Markovic}, {Marsh}, {Marulli}, {Massey}, {Mellier}, {Montanari}, {Mota}, {Nunes}, {Percival}, {Pettorino}, {Porciani}, {Quercellini}, {Read}, {Rinaldi}, {Sapone}, {Sawicki}, {Scaramella}, {Skordis}, {Simpson}, {Taylor}, {Thomas}, {Trotta}, {Verde}, {Vernizzi}, {Vollmer}, {Wang}, {Weller}, \& {Zlosnik}}]{Amendola2018}
{Amendola}, L., {Appleby}, S., {Avgoustidis}, A., {et~al.} 2018, Living Reviews in Relativity, 21, 2

\bibitem[{{Astropy Collaboration} {et~al.}(2022){Astropy Collaboration}, {Price-Whelan}, {Lim}, {Earl}, {Starkman}, {Bradley}, {Shupe}, {Patil}, {Corrales}, {Brasseur}, {N{"o}the}, {Donath}, {Tollerud}, {Morris}, {Ginsburg}, {Vaher}, {Weaver}, {Tocknell}, {Jamieson}, {van Kerkwijk}, {Robitaille}, {Merry}, {Bachetti}, {G{"u}nther}, {Aldcroft}, {Alvarado-Montes}, {Archibald}, {B{'o}di}, {Bapat}, {Barentsen}, {Baz{'a}n}, {Biswas}, {Boquien}, {Burke}, {Cara}, {Cara}, {Conroy}, {Conseil}, {Craig}, {Cross}, {Cruz}, {D'Eugenio}, {Dencheva}, {Devillepoix}, {Dietrich}, {Eigenbrot}, {Erben}, {Ferreira}, {Foreman-Mackey}, {Fox}, {Freij}, {Garg}, {Geda}, {Glattly}, {Gondhalekar}, {Gordon}, {Grant}, {Greenfield}, {Groener}, {Guest}, {Gurovich}, {Handberg}, {Hart}, {Hatfield-Dodds}, {Homeier}, {Hosseinzadeh}, {Jenness}, {Jones}, {Joseph}, {Kalmbach}, {Karamehmetoglu}, {Ka{l}uszy{'n}ski}, {Kelley}, {Kern}, {Kerzendorf}, {Koch}, {Kulumani}, {Lee}, {Ly}, {Ma}, {MacBride}, {Maljaars}, {Muna}, {Murphy}, {Norman}, {O'Steen},
  {Oman}, {Pacifici}, {Pascual}, {Pascual-Granado}, {Patil}, {Perren}, {Pickering}, {Rastogi}, {Roulston}, {Ryan}, {Rykoff}, {Sabater}, {Sakurikar}, {Salgado}, {Sanghi}, {Saunders}, {Savchenko}, {Schwardt}, {Seifert-Eckert}, {Shih}, {Jain}, {Shukla}, {Sick}, {Simpson}, {Singanamalla}, {Singer}, {Singhal}, {Sinha}, {Sip{H{o}}cz}, {Spitler}, {Stansby}, {Streicher}, {{{S}}umak}, {Swinbank}, {Taranu}, {Tewary}, {Tremblay}, {Val-Borro}, {Van Kooten}, {Vasovi{'c}}, {Verma}, {de Miranda Cardoso}, {Williams}, {Wilson}, {Winkel}, {Wood-Vasey}, {Xue}, {Yoachim}, {Zhang}, {Zonca}, \& {Astropy Project Contributors}}]{astropy2022}
{Astropy Collaboration}, {Price-Whelan}, A.~M., {Lim}, P.~L., {et~al.} 2022, \apj, 935, 167

\bibitem[{{Astropy Collaboration} {et~al.}(2018){Astropy Collaboration}, {Price-Whelan}, {Sip{\H{o}}cz}, {G{\"u}nther}, {Lim}, {Crawford}, {Conseil}, {Shupe}, {Craig}, {Dencheva}, {Ginsburg}, {Vand erPlas}, {Bradley}, {P{\'e}rez-Su{\'a}rez}, {de Val-Borro}, {Aldcroft}, {Cruz}, {Robitaille}, {Tollerud}, {Ardelean}, {Babej}, {Bach}, {Bachetti}, {Bakanov}, {Bamford}, {Barentsen}, {Barmby}, {Baumbach}, {Berry}, {Biscani}, {Boquien}, {Bostroem}, {Bouma}, {Brammer}, {Bray}, {Breytenbach}, {Buddelmeijer}, {Burke}, {Calderone}, {Cano Rodr{\'\i}guez}, {Cara}, {Cardoso}, {Cheedella}, {Copin}, {Corrales}, {Crichton}, {D'Avella}, {Deil}, {Depagne}, {Dietrich}, {Donath}, {Droettboom}, {Earl}, {Erben}, {Fabbro}, {Ferreira}, {Finethy}, {Fox}, {Garrison}, {Gibbons}, {Goldstein}, {Gommers}, {Greco}, {Greenfield}, {Groener}, {Grollier}, {Hagen}, {Hirst}, {Homeier}, {Horton}, {Hosseinzadeh}, {Hu}, {Hunkeler}, {Ivezi{\'c}}, {Jain}, {Jenness}, {Kanarek}, {Kendrew}, {Kern}, {Kerzendorf}, {Khvalko}, {King}, {Kirkby}, {Kulkarni},
  {Kumar}, {Lee}, {Lenz}, {Littlefair}, {Ma}, {Macleod}, {Mastropietro}, {McCully}, {Montagnac}, {Morris}, {Mueller}, {Mumford}, {Muna}, {Murphy}, {Nelson}, {Nguyen}, {Ninan}, {N{\"o}the}, {Ogaz}, {Oh}, {Parejko}, {Parley}, {Pascual}, {Patil}, {Patil}, {Plunkett}, {Prochaska}, {Rastogi}, {Reddy Janga}, {Sabater}, {Sakurikar}, {Seifert}, {Sherbert}, {Sherwood-Taylor}, {Shih}, {Sick}, {Silbiger}, {Singanamalla}, {Singer}, {Sladen}, {Sooley}, {Sornarajah}, {Streicher}, {Teuben}, {Thomas}, {Tremblay}, {Turner}, {Terr{\'o}n}, {van Kerkwijk}, {de la Vega}, {Watkins}, {Weaver}, {Whitmore}, {Woillez}, {Zabalza}, \& {Astropy Contributors}}]{astropy2018}
{Astropy Collaboration}, {Price-Whelan}, A.~M., {Sip{\H{o}}cz}, B.~M., {et~al.} 2018, \aj, 156, 123

\bibitem[{{Astropy Collaboration} {et~al.}(2013){Astropy Collaboration}, {Robitaille}, {Tollerud}, {Greenfield}, {Droettboom}, {Bray}, {Aldcroft}, {Davis}, {Ginsburg}, {Price-Whelan}, {Kerzendorf}, {Conley}, {Crighton}, {Barbary}, {Muna}, {Ferguson}, {Grollier}, {Parikh}, {Nair}, {Unther}, {Deil}, {Woillez}, {Conseil}, {Kramer}, {Turner}, {Singer}, {Fox}, {Weaver}, {Zabalza}, {Edwards}, {Azalee Bostroem}, {Burke}, {Casey}, {Crawford}, {Dencheva}, {Ely}, {Jenness}, {Labrie}, {Lim}, {Pierfederici}, {Pontzen}, {Ptak}, {Refsdal}, {Servillat}, \& {Streicher}}]{astropy2013}
{Astropy Collaboration}, {Robitaille}, T.~P., {Tollerud}, E.~J., {et~al.} 2013, \aap, 558, A33

\bibitem[{{Bartelmann} \& {Schneider}(2001)}]{Bartelmann2001}
{Bartelmann}, M. \& {Schneider}, P. 2001, \physrep, 340, 291

\bibitem[{{Battye} {et~al.}(2020){Battye}, {Brown}, {Casey}, {Harrison}, {Jackson}, {Smail}, {Watson}, {Hales}, {Manning}, {Hung}, {Riseley}, {Abdalla}, {Birkinshaw}, {Demetroullas}, {Chapman}, {Beswick}, {Muxlow}, {Bonaldi}, {Camera}, {Hillier}, {Kay}, {Peters}, {Sanders}, {Thomas}, {Thomson}, {Tunbridge}, {Whittaker}, \& {SuperCLASS Collaboration}}]{Battye2020}
{Battye}, R.~A., {Brown}, M.~L., {Casey}, C.~M., {et~al.} 2020, \mnras, 495, 1706

\bibitem[{{Becker} {et~al.}(1995){Becker}, {White}, \& {Helfand}}]{Becker1995}
{Becker}, R.~H., {White}, R.~L., \& {Helfand}, D.~J. 1995, \apj, 450, 559

\bibitem[{{Bernstein} \& {Jarvis}(2002)}]{Bernstein2002}
{Bernstein}, G.~M. \& {Jarvis}, M. 2002, \aj, 123, 583

\bibitem[{{Bertin} \& {Arnouts}(1996)}]{Bertin1996}
{Bertin}, E. \& {Arnouts}, S. 1996, \aaps, 117, 393

\bibitem[{{Best} {et~al.}(2023){Best}, {Kondapally}, {Williams}, {Cochrane}, {Duncan}, {Hale}, {Haskell}, {Ma{\l}ek}, {McCheyne}, {Smith}, {Wang}, {Botteon}, {Bonato}, {Bondi}, {Calistro Rivera}, {Gao}, {G{\"u}rkan}, {Hardcastle}, {Jarvis}, {Mingo}, {Miraghaei}, {Morabito}, {Nisbet}, {Prandoni}, {R{\"o}ttgering}, {Sabater}, {Shimwell}, {Tasse}, \& {van Weeren}}]{Best2023}
{Best}, P.~N., {Kondapally}, R., {Williams}, W.~L., {et~al.} 2023, \mnras, 523, 1729

\bibitem[{Bonaldi {et~al.}(2018)Bonaldi, Bonato, Galluzzi, Harrison, Massardi, Kay, De~Zotti, \& Brown}]{Bonaldi2018}
Bonaldi, A., Bonato, M., Galluzzi, V., {et~al.} 2018, Monthly Notices of the Royal Astronomical Society, 482, 2

\bibitem[{{Bonaldi} {et~al.}(2016){Bonaldi}, {Harrison}, {Camera}, \& {Brown}}]{Bonaldi2016}
{Bonaldi}, A., {Harrison}, I., {Camera}, S., \& {Brown}, M.~L. 2016, \mnras, 463, 3686

\bibitem[{{Bosch} {et~al.}(2018){Bosch}, {Armstrong}, {Bickerton}, {Furusawa}, {Ikeda}, {Koike}, {Lupton}, {Mineo}, {Price}, {Takata}, {Tanaka}, {Yasuda}, {AlSayyad}, {Becker}, {Coulton}, {Coupon}, {Garmilla}, {Huang}, {Krughoff}, {Lang}, {Leauthaud}, {Lim}, {Lust}, {MacArthur}, {Mandelbaum}, {Miyatake}, {Miyazaki}, {Murata}, {More}, {Okura}, {Owen}, {Swinbank}, {Strauss}, {Yamada}, \& {Yamanoi}}]{Bosch2018}
{Bosch}, J., {Armstrong}, R., {Bickerton}, S., {et~al.} 2018, \pasj, 70, S5

\bibitem[{{Brown} {et~al.}(2015){Brown}, {Bacon}, {Camera}, {Harrison}, {Joachimi}, {Metcalf}, {Pourtsidou}, {Takahashi}, {Zuntz}, {Abdalla}, {Bridle}, {Jarvis}, {Kitching}, {Miller}, \& {Patel}}]{Brown2015}
{Brown}, M., {Bacon}, D., {Camera}, S., {et~al.} 2015, in Advancing Astrophysics with the Square Kilometre Array (AASKA14), 23

\bibitem[{{Brown} \& {Battye}(2011)}]{Brown2011}
{Brown}, M.~L. \& {Battye}, R.~A. 2011, \mnras, 410, 2057

\bibitem[{{Camera} {et~al.}(2017){Camera}, {Harrison}, {Bonaldi}, \& {Brown}}]{Camera2017}
{Camera}, S., {Harrison}, I., {Bonaldi}, A., \& {Brown}, M.~L. 2017, \mnras, 464, 4747

\bibitem[{{Carrillo} {et~al.}(2012){Carrillo}, {McEwen}, \& {Wiaux}}]{Carrillo2012}
{Carrillo}, R.~E., {McEwen}, J.~D., \& {Wiaux}, Y. 2012, \mnras, 426, 1223

\bibitem[{{Carrillo} {et~al.}(2014){Carrillo}, {McEwen}, \& {Wiaux}}]{Carrillo2014}
{Carrillo}, R.~E., {McEwen}, J.~D., \& {Wiaux}, Y. 2014, \mnras, 439, 3591

\bibitem[{{Chang} {et~al.}(2004){Chang}, {Refregier}, \& {Helfand}}]{Chang2004}
{Chang}, T.-C., {Refregier}, A., \& {Helfand}, D.~J. 2004, \apj, 617, 794

\bibitem[{{Chisari} {et~al.}(2019){Chisari}, {Alonso}, {Krause}, {Leonard}, {Bull}, {Neveu}, {Villarreal}, {Singh}, {McClintock}, {Ellison}, {Du}, {Zuntz}, {Mead}, {Joudaki}, {Lorenz}, {Tr{\"o}ster}, {Sanchez}, {Lanusse}, {Ishak}, {Hlozek}, {Blazek}, {Campagne}, {Almoubayyed}, {Eifler}, {Kirby}, {Kirkby}, {Plaszczynski}, {Slosar}, {Vrastil}, {Wagoner}, \& {LSST Dark Energy Science Collaboration}}]{Chisari2019}
{Chisari}, N.~E., {Alonso}, D., {Krause}, E., {et~al.} 2019, \apjs, 242, 2

\bibitem[{{Connor} {et~al.}(2022){Connor}, {Bouman}, {Ravi}, \& {Hallinan}}]{Connor2022}
{Connor}, L., {Bouman}, K.~L., {Ravi}, V., \& {Hallinan}, G. 2022, \mnras, 514, 2614

\bibitem[{{Dabbech} {et~al.}(2015){Dabbech}, {Ferrari}, {Mary}, {Slezak}, {Smirnov}, \& {Kenyon}}]{Dabbech2015}
{Dabbech}, A., {Ferrari}, C., {Mary}, D., {et~al.} 2015, \aap, 576, A7

\bibitem[{{Dalal} {et~al.}(2023){Dalal}, {Li}, {Nicola}, {Zuntz}, {Strauss}, {Sugiyama}, {Zhang}, {Rau}, {Mandelbaum}, {Takada}, {More}, {Miyatake}, {Kannawadi}, {Shirasaki}, {Taniguchi}, {Takahashi}, {Osato}, {Hamana}, {Oguri}, {Nishizawa}, {Malag{\'o}n}, {Sunayama}, {Alonso}, {Slosar}, {Luo}, {Armstrong}, {Bosch}, {Hsieh}, {Komiyama}, {Lupton}, {Lust}, {MacArthur}, {Miyazaki}, {Murayama}, {Nishimichi}, {Okura}, {Price}, {Tait}, {Tanaka}, \& {Wang}}]{Dalal2023}
{Dalal}, R., {Li}, X., {Nicola}, A., {et~al.} 2023, \prd, 108, 123519

\bibitem[{{de Jong} {et~al.}(2024){de Jong}, {van Weeren}, {Sweijen}, {Oonk}, {Shimwell}, {Offringa}, {Morabito}, {R{\"o}ttgering}, {Kondapally}, {Escott}, {Best}, {Bondi}, {Ye}, \& {Petley}}]{deJong2024}
{de Jong}, J.~M.~G.~H.~J., {van Weeren}, R.~J., {Sweijen}, F., {et~al.} 2024, arXiv e-prints, arXiv:2407.13247

\bibitem[{{de Jong} {et~al.}(2013){de Jong}, {Verdoes Kleijn}, {Kuijken}, \& {Valentijn}}]{deJong2013}
{de Jong}, J. T.~A., {Verdoes Kleijn}, G.~A., {Kuijken}, K.~H., \& {Valentijn}, E.~A. 2013, Experimental Astronomy, 35, 25

\bibitem[{{Demetroullas} \& {Brown}(2016)}]{Demetroullas2016}
{Demetroullas}, C. \& {Brown}, M.~L. 2016, \mnras, 456, 3100

\bibitem[{{Demetroullas} \& {Brown}(2018)}]{Demetroullas2018}
{Demetroullas}, C. \& {Brown}, M.~L. 2018, \mnras, 473, 937

\bibitem[{{Euclid Collaboration} {et~al.}(2024){Euclid Collaboration}, {Mellier}, {Abdurro'uf}, {Acevedo Barroso}, {Ach{\'u}carro}, {Adamek}, {Adam}, {Addison}, {Aghanim}, {Aguena}, \& et~al.}]{EuclidCollaboration2024}
{Euclid Collaboration}, {Mellier}, Y., {Abdurro'uf}, {et~al.} 2024, arXiv e-prints, arXiv:2405.13491

\bibitem[{{Fernique} {et~al.}(2014){Fernique}, {Boch}, {Donaldson}, {Durand}, {O'Mullane}, {Reinecke}, \& {Taylor}}]{Fernique2014}
{Fernique}, P., {Boch}, T., {Donaldson}, T., {et~al.} 2014, {MOC - HEALPix Multi-Order Coverage map Version 1.0}, IVOA Recommendation 02 June 2014

\bibitem[{{Fu} {et~al.}(2008){Fu}, {Semboloni}, {Hoekstra}, {Kilbinger}, {van Waerbeke}, {Tereno}, {Mellier}, {Heymans}, {Coupon}, {Benabed}, {Benjamin}, {Bertin}, {Dor{\'e}}, {Hudson}, {Ilbert}, {Maoli}, {Marmo}, {McCracken}, \& {M{\'e}nard}}]{Fu2008}
{Fu}, L., {Semboloni}, E., {Hoekstra}, H., {et~al.} 2008, \aap, 479, 9

\bibitem[{{Gatti} {et~al.}(2021){Gatti}, {Sheldon}, {Amon}, {Becker}, {Troxel}, {Choi}, {Doux}, {MacCrann}, {Navarro-Alsina}, {Harrison}, {Gruen}, {Bernstein}, {Jarvis}, {Secco}, {Fert{\'e}}, {Shin}, {McCullough}, {Rollins}, {Chen}, {Chang}, {Pandey}, {Tutusaus}, {Prat}, {Elvin-Poole}, {Sanchez}, {Plazas}, {Roodman}, {Zuntz}, {Abbott}, {Aguena}, {Allam}, {Annis}, {Avila}, {Bacon}, {Bertin}, {Bhargava}, {Brooks}, {Burke}, {Carnero Rosell}, {Carrasco Kind}, {Carretero}, {Castander}, {Conselice}, {Costanzi}, {Crocce}, {da Costa}, {Davis}, {De Vicente}, {Desai}, {Diehl}, {Dietrich}, {Doel}, {Drlica-Wagner}, {Eckert}, {Everett}, {Ferrero}, {Frieman}, {Garc{\'\i}a-Bellido}, {Gerdes}, {Giannantonio}, {Gruendl}, {Gschwend}, {Gutierrez}, {Hartley}, {Hinton}, {Hollowood}, {Honscheid}, {Hoyle}, {Huff}, {Huterer}, {Jain}, {James}, {Jeltema}, {Krause}, {Kron}, {Kuropatkin}, {Lima}, {Maia}, {Marshall}, {Miquel}, {Morgan}, {Myles}, {Palmese}, {Paz-Chinch{\'o}n}, {Rykoff}, {Samuroff}, {Sanchez}, {Scarpine}, {Schubnell},
  {Serrano}, {Sevilla-Noarbe}, {Smith}, {Soares-Santos}, {Suchyta}, {Swanson}, {Tarle}, {Thomas}, {To}, {Tucker}, {Varga}, {Wechsler}, {Weller}, {Wester}, \& {Wilkinson}}]{Gatti2021}
{Gatti}, M., {Sheldon}, E., {Amon}, A., {et~al.} 2021, \mnras, 504, 4312

\bibitem[{{Giblin} {et~al.}(2021){Giblin}, {Heymans}, {Asgari}, {Hildebrandt}, {Hoekstra}, {Joachimi}, {Kannawadi}, {Kuijken}, {Lin}, {Miller}, {Tr{\"o}ster}, {van den Busch}, {Wright}, {Bilicki}, {Blake}, {de Jong}, {Dvornik}, {Erben}, {Getman}, {Napolitano}, {Schneider}, {Shan}, \& {Valentijn}}]{Giblin2021}
{Giblin}, B., {Heymans}, C., {Asgari}, M., {et~al.} 2021, \aap, 645, A105

\bibitem[{{G{\'o}rski} {et~al.}(2005){G{\'o}rski}, {Hivon}, {Banday}, {Wandelt}, {Hansen}, {Reinecke}, \& {Bartelmann}}]{Gorski2005}
{G{\'o}rski}, K.~M., {Hivon}, E., {Banday}, A.~J., {et~al.} 2005, \apj, 622, 759

\bibitem[{{Hamana} {et~al.}(2020){Hamana}, {Shirasaki}, {Miyazaki}, {Hikage}, {Oguri}, {More}, {Armstrong}, {Leauthaud}, {Mandelbaum}, {Miyatake}, {Nishizawa}, {Simet}, {Takada}, {Aihara}, {Bosch}, {Komiyama}, {Lupton}, {Murayama}, {Strauss}, \& {Tanaka}}]{Hamana2020}
{Hamana}, T., {Shirasaki}, M., {Miyazaki}, S., {et~al.} 2020, \pasj, 72, 16

\bibitem[{Harris {et~al.}(2020)Harris, Millman, van~der Walt, Gommers, Virtanen, Cournapeau, Wieser, Taylor, Berg, Smith, Kern, Picus, Hoyer, van Kerkwijk, Brett, Haldane, del R{\'{i}}o, Wiebe, Peterson, G{\'{e}}rard-Marchant, Sheppard, Reddy, Weckesser, Abbasi, Gohlke, \& Oliphant}]{harris2020array}
Harris, C.~R., Millman, K.~J., van~der Walt, S.~J., {et~al.} 2020, Nature, 585, 357

\bibitem[{{Harrison} {et~al.}(2020){Harrison}, {Brown}, {Tunbridge}, {Thomas}, {Hillier}, {Thomson}, {Whittaker}, {Abdalla}, {Battye}, {Bonaldi}, {Camera}, {Casey}, {Demetroullas}, {Hales}, {Jackson}, {Kay}, {Manning}, {Peters}, {Riseley}, {Watson}, \& {SuperCLASS Collaboration}}]{Harrison2020}
{Harrison}, I., {Brown}, M.~L., {Tunbridge}, B., {et~al.} 2020, \mnras, 495, 1737

\bibitem[{{Harrison} {et~al.}(2016){Harrison}, {Camera}, {Zuntz}, \& {Brown}}]{Harrison2016}
{Harrison}, I., {Camera}, S., {Zuntz}, J., \& {Brown}, M.~L. 2016, \mnras, 463, 3674

\bibitem[{{Heymans} {et~al.}(2012){Heymans}, {Van Waerbeke}, {Miller}, {Erben}, {Hildebrandt}, {Hoekstra}, {Kitching}, {Mellier}, {Simon}, {Bonnett}, {Coupon}, {Fu}, {Harnois D{\'e}raps}, {Hudson}, {Kilbinger}, {Kuijken}, {Rowe}, {Schrabback}, {Semboloni}, {van Uitert}, {Vafaei}, \& {Velander}}]{Heymans2012}
{Heymans}, C., {Van Waerbeke}, L., {Miller}, L., {et~al.} 2012, \mnras, 427, 146

\bibitem[{{Hikage} {et~al.}(2019){Hikage}, {Oguri}, {Hamana}, {More}, {Mandelbaum}, {Takada}, {K{\"o}hlinger}, {Miyatake}, {Nishizawa}, {Aihara}, {Armstrong}, {Bosch}, {Coupon}, {Ducout}, {Ho}, {Hsieh}, {Komiyama}, {Lanusse}, {Leauthaud}, {Lupton}, {Medezinski}, {Mineo}, {Miyama}, {Miyazaki}, {Murata}, {Murayama}, {Shirasaki}, {Sif{\'o}n}, {Simet}, {Speagle}, {Spergel}, {Strauss}, {Sugiyama}, {Tanaka}, {Utsumi}, {Wang}, \& {Yamada}}]{Hikage2019}
{Hikage}, C., {Oguri}, M., {Hamana}, T., {et~al.} 2019, \pasj, 71, 43

\bibitem[{{Hillier} {et~al.}(2019){Hillier}, {Brown}, {Harrison}, \& {Whittaker}}]{Hillier2019}
{Hillier}, T., {Brown}, M.~L., {Harrison}, I., \& {Whittaker}, L. 2019, \mnras, 488, 5420

\bibitem[{{Hirata} \& {Seljak}(2003)}]{Hirata2003}
{Hirata}, C. \& {Seljak}, U. 2003, \mnras, 343, 459

\bibitem[{{H{\"o}gbom}(1974)}]{Hogbom1974}
{H{\"o}gbom}, J.~A. 1974, \aaps, 15, 417

\bibitem[{Hunter(2007)}]{Hunter2007Matplotlib}
Hunter, J.~D. 2007, Computing in Science \& Engineering, 9, 90

\bibitem[{{Ivezi{\'c}} {et~al.}(2019){Ivezi{\'c}}, {Kahn}, {Tyson}, {Abel}, {Acosta}, {Allsman}, {Alonso}, {AlSayyad}, {Anderson}, {Andrew}, \& et~al.}]{Ivezic2019}
{Ivezi{\'c}}, {\v{Z}}., {Kahn}, S.~M., {Tyson}, J.~A., {et~al.} 2019, \apj, 873, 111

\bibitem[{{Jarvis} {et~al.}(2004){Jarvis}, {Bernstein}, \& {Jain}}]{Jarvis2004}
{Jarvis}, M., {Bernstein}, G., \& {Jain}, B. 2004, \mnras, 352, 338

\bibitem[{{Junklewitz} {et~al.}(2016){Junklewitz}, {Bell}, {Selig}, \& {En{\ss}lin}}]{Junklewitz2016}
{Junklewitz}, H., {Bell}, M.~R., {Selig}, M., \& {En{\ss}lin}, T.~A. 2016, \aap, 586, A76

\bibitem[{{Kaiser} {et~al.}(1995){Kaiser}, {Squires}, \& {Broadhurst}}]{Kaiser1995}
{Kaiser}, N., {Squires}, G., \& {Broadhurst}, T. 1995, \apj, 449, 460

\bibitem[{{Kilbinger}(2015)}]{Kilbinger2015}
{Kilbinger}, M. 2015, Reports on Progress in Physics, 78, 086901

\bibitem[{{Kondapally} {et~al.}(2021){Kondapally}, {Best}, {Hardcastle}, {Nisbet}, {Bonato}, {Sabater}, {Duncan}, {McCheyne}, {Cochrane}, {Bowler}, {Williams}, {Shimwell}, {Tasse}, {Croston}, {Goyal}, {Jamrozy}, {Jarvis}, {Mahatma}, {R{\"o}ttgering}, {Smith}, {Wo{\l}owska}, {Bondi}, {Brienza}, {Brown}, {Br{\"u}ggen}, {Chambers}, {Garrett}, {G{\"u}rkan}, {Huber}, {Kunert-Bajraszewska}, {Magnier}, {Mingo}, {Mostert}, {Nikiel-Wroczy{\'n}ski}, {O'Sullivan}, {Paladino}, {Ploeckinger}, {Prandoni}, {Rosenthal}, {Schwarz}, {Shulevski}, {Wagenveld}, \& {Wang}}]{Kondapally2021}
{Kondapally}, R., {Best}, P.~N., {Hardcastle}, M.~J., {et~al.} 2021, \aap, 648, A3

\bibitem[{{Kuijken} {et~al.}(2015){Kuijken}, {Heymans}, {Hildebrandt}, {Nakajima}, {Erben}, {de Jong}, {Viola}, {Choi}, {Hoekstra}, {Miller}, {van Uitert}, {Amon}, {Blake}, {Brouwer}, {Buddendiek}, {Conti}, {Eriksen}, {Grado}, {Harnois-D{\'e}raps}, {Helmich}, {Herbonnet}, {Irisarri}, {Kitching}, {Klaes}, {La Barbera}, {Napolitano}, {Radovich}, {Schneider}, {Sif{\'o}n}, {Sikkema}, {Simon}, {Tudorica}, {Valentijn}, {Verdoes Kleijn}, \& {van Waerbeke}}]{Kuijken2015}
{Kuijken}, K., {Heymans}, C., {Hildebrandt}, H., {et~al.} 2015, \mnras, 454, 3500

\bibitem[{{Laureijs} {et~al.}(2011){Laureijs}, {Amiaux}, {Arduini}, {Augu{\`e}res}, {Brinchmann}, {Cole}, {Cropper}, {Dabin}, {Duvet}, {Ealet}, \& et~al.}]{Laureijs2011}
{Laureijs}, R., {Amiaux}, J., {Arduini}, S., {et~al.} 2011, arXiv e-prints, arXiv:1110.3193

\bibitem[{{Limber}(1953)}]{Limber1953}
{Limber}, D.~N. 1953, \apj, 117, 134

\bibitem[{{Mandelbaum} {et~al.}(2018{\natexlab{a}}){Mandelbaum}, {Lanusse}, {Leauthaud}, {Armstrong}, {Simet}, {Miyatake}, {Meyers}, {Bosch}, {Murata}, {Miyazaki}, \& {Tanaka}}]{Mandelbaum2018Dec}
{Mandelbaum}, R., {Lanusse}, F., {Leauthaud}, A., {et~al.} 2018{\natexlab{a}}, \mnras, 481, 3170

\bibitem[{{Mandelbaum} {et~al.}(2018{\natexlab{b}}){Mandelbaum}, {Miyatake}, {Hamana}, {Oguri}, {Simet}, {Armstrong}, {Bosch}, {Murata}, {Lanusse}, {Leauthaud}, {Coupon}, {More}, {Takada}, {Miyazaki}, {Speagle}, {Shirasaki}, {Sif{\'o}n}, {Huang}, {Nishizawa}, {Medezinski}, {Okura}, {Okabe}, {Czakon}, {Takahashi}, {Coulton}, {Hikage}, {Komiyama}, {Lupton}, {Strauss}, {Tanaka}, \& {Utsumi}}]{Mandelbaum2018Jan}
{Mandelbaum}, R., {Miyatake}, H., {Hamana}, T., {et~al.} 2018{\natexlab{b}}, \pasj, 70, S25

\bibitem[{{Mohan} \& {Rafferty}(2015)}]{Mohan2015}
{Mohan}, N. \& {Rafferty}, D. 2015, {PyBDSF: Python Blob Detection and Source Finder}, Astrophysics Source Code Library, record ascl:1502.007

\bibitem[{{Morabito} {et~al.}(2022){Morabito}, {Sweijen}, {Radcliffe}, {Best}, {Kondapally}, {Bondi}, {Bonato}, {Duncan}, {Prandoni}, {Shimwell}, {Williams}, {van Weeren}, {Conway}, \& {Calistro Rivera}}]{Morabito2022}
{Morabito}, L.~K., {Sweijen}, F., {Radcliffe}, J.~F., {et~al.} 2022, \mnras, 515, 5758

\bibitem[{{Oliver} {et~al.}(2000){Oliver}, {Rowan-Robinson}, {Alexander}, {Almaini}, {Balcells}, {Baker}, {Barcons}, {Barden}, {Bellas-Velidis}, {Cabrera-Guerra}, {Carballo}, {Cesarsky}, {Ciliegi}, {Clements}, {Crockett}, {Danese}, {Dapergolas}, {Drolias}, {Eaton}, {Efstathiou}, {Egami}, {Elbaz}, {Fadda}, {Fox}, {Franceschini}, {Genzel}, {Goldschmidt}, {Graham}, {Gonzalez-Serrano}, {Gonzalez-Solares}, {Granato}, {Gruppioni}, {Herbstmeier}, {H{\'e}raudeau}, {Joshi}, {Kontizas}, {Kontizas}, {Kotilainen}, {Kunze}, {La Franca}, {Lari}, {Lawrence}, {Lemke}, {Linden-V{\o}rnle}, {Mann}, {M{\'a}rquez}, {Masegosa}, {Mattila}, {McMahon}, {Miley}, {Missoulis}, {Mobasher}, {Morel}, {N{\o}rgaard-Nielsen}, {Omont}, {Papadopoulos}, {Perez-Fournon}, {Puget}, {Rigopoulou}, {Rocca-Volmerange}, {Serjeant}, {Silva}, {Sumner}, {Surace}, {Vaisanen}, {van der Werf}, {Verma}, {Vigroux}, {Villar-Martin}, \& {Willott}}]{Oliver2000}
{Oliver}, S., {Rowan-Robinson}, M., {Alexander}, D.~M., {et~al.} 2000, \mnras, 316, 749

\bibitem[{pandas~development team(2020)}]{pandas2020}
pandas~development team, T. 2020, pandas-dev/pandas: Pandas

\bibitem[{{Patel} {et~al.}(2010){Patel}, {Bacon}, {Beswick}, {Muxlow}, \& {Hoyle}}]{Patel2010}
{Patel}, P., {Bacon}, D.~J., {Beswick}, R.~J., {Muxlow}, T.~W.~B., \& {Hoyle}, B. 2010, \mnras, 401, 2572

\bibitem[{{Rivi} \& {Miller}(2018)}]{Rivi2018}
{Rivi}, M. \& {Miller}, L. 2018, \mnras, 476, 2053

\bibitem[{{Rivi} {et~al.}(2016){Rivi}, {Miller}, {Makhathini}, \& {Abdalla}}]{Rivi2016}
{Rivi}, M., {Miller}, L., {Makhathini}, S., \& {Abdalla}, F.~B. 2016, \mnras, 463, 1881

\bibitem[{{Sabater} {et~al.}(2021){Sabater}, {Best}, {Tasse}, {Hardcastle}, {Shimwell}, {Nisbet}, {Jelic}, {Callingham}, {R{\"o}ttgering}, {Bonato}, {Bondi}, {Ciardi}, {Cochrane}, {Jarvis}, {Kondapally}, {Koopmans}, {O'Sullivan}, {Prandoni}, {Schwarz}, {Smith}, {Wang}, {Williams}, \& {Zaroubi}}]{Sabater2021}
{Sabater}, J., {Best}, P.~N., {Tasse}, C., {et~al.} 2021, \aap, 648, A2

\bibitem[{{Shimwell} {et~al.}(2025){Shimwell}, {Hale}, {Best}, {Botteon}, {Drabent}, {Hardcastle}, {Jeli{\'c}}, {de Jong}, {Kondapally}, {R{\"o}ttgering}, {Tasse}, {van Weeren}, {Williams}, {Bonafede}, {Bondi}, {Br{\"u}ggen}, {Brunetti}, {Callingham}, {De Gasperin}, {Duncan}, {Horellou}, {Iyer}, {de Ruiter}, {Ma{\l}ek}, {Nair}, {Morabito}, {Prandoni}, {Rowlinson}, {Sabater}, {Shulevski}, {Smith}, \& {Sweijen}}]{Shimwell2024}
{Shimwell}, T.~W., {Hale}, C.~L., {Best}, P.~N., {et~al.} 2025, arXiv e-prints, arXiv:2501.04093

\bibitem[{{Shimwell} {et~al.}(2022){Shimwell}, {Hardcastle}, {Tasse}, {Best}, {R{\"o}ttgering}, {Williams}, {Botteon}, {Drabent}, {Mechev}, {Shulevski}, {van Weeren}, {Bester}, {Br{\"u}ggen}, {Brunetti}, {Callingham}, {Chy{\.z}y}, {Conway}, {Dijkema}, {Duncan}, {de Gasperin}, {Hale}, {Haverkorn}, {Hugo}, {Jackson}, {Mevius}, {Miley}, {Morabito}, {Morganti}, {Offringa}, {Oonk}, {Rafferty}, {Sabater}, {Smith}, {Schwarz}, {Smirnov}, {O'Sullivan}, {Vedantham}, {White}, {Albert}, {Alegre}, {Asabere}, {Bacon}, {Bonafede}, {Bonnassieux}, {Brienza}, {Bilicki}, {Bonato}, {Calistro Rivera}, {Cassano}, {Cochrane}, {Croston}, {Cuciti}, {Dallacasa}, {Danezi}, {Dettmar}, {Di Gennaro}, {Edler}, {En{\ss}lin}, {Emig}, {Franzen}, {Garc{\'\i}a-Vergara}, {Grange}, {G{\"u}rkan}, {Hajduk}, {Heald}, {Heesen}, {Hoang}, {Hoeft}, {Horellou}, {Iacobelli}, {Jamrozy}, {Jeli{\'c}}, {Kondapally}, {Kukreti}, {Kunert-Bajraszewska}, {Magliocchetti}, {Mahatma}, {Ma{\l}ek}, {Mandal}, {Massaro}, {Meyer-Zhao}, {Mingo}, {Mostert}, {Nair},
  {Nakoneczny}, {Nikiel-Wroczy{\'n}ski}, {Orr{\'u}}, {Pajdosz-{\'S}mierciak}, {Pasini}, {Prandoni}, {van Piggelen}, {Rajpurohit}, {Retana-Montenegro}, {Riseley}, {Rowlinson}, {Saxena}, {Schrijvers}, {Sweijen}, {Siewert}, {Timmerman}, {Vaccari}, {Vink}, {West}, {Wo{\l}owska}, {Zhang}, \& {Zheng}}]{Shimwell2022}
{Shimwell}, T.~W., {Hardcastle}, M.~J., {Tasse}, C., {et~al.} 2022, \aap, 659, A1

\bibitem[{{Shimwell} {et~al.}(2017){Shimwell}, {R{\"o}ttgering}, {Best}, {Williams}, {Dijkema}, {de Gasperin}, {Hardcastle}, {Heald}, {Hoang}, {Horneffer}, {Intema}, {Mahony}, {Mandal}, {Mechev}, {Morabito}, {Oonk}, {Rafferty}, {Retana-Montenegro}, {Sabater}, {Tasse}, {van Weeren}, {Br{\"u}ggen}, {Brunetti}, {Chy{\.z}y}, {Conway}, {Haverkorn}, {Jackson}, {Jarvis}, {McKean}, {Miley}, {Morganti}, {White}, {Wise}, {van Bemmel}, {Beck}, {Brienza}, {Bonafede}, {Calistro Rivera}, {Cassano}, {Clarke}, {Cseh}, {Deller}, {Drabent}, {van Driel}, {Engels}, {Falcke}, {Ferrari}, {Fr{\"o}hlich}, {Garrett}, {Harwood}, {Heesen}, {Hoeft}, {Horellou}, {Israel}, {Kapi{\'n}ska}, {Kunert-Bajraszewska}, {McKay}, {Mohan}, {Orr{\'u}}, {Pizzo}, {Prandoni}, {Schwarz}, {Shulevski}, {Sipior}, {Smith}, {Sridhar}, {Steinmetz}, {Stroe}, {Varenius}, {van der Werf}, {Zensus}, \& {Zwart}}]{Shimwell2017}
{Shimwell}, T.~W., {R{\"o}ttgering}, H.~J.~A., {Best}, P.~N., {et~al.} 2017, \aap, 598, A104

\bibitem[{{Shimwell} {et~al.}(2019){Shimwell}, {Tasse}, {Hardcastle}, {Mechev}, {Williams}, {Best}, {R{\"o}ttgering}, {Callingham}, {Dijkema}, {de Gasperin}, {Hoang}, {Hugo}, {Mirmont}, {Oonk}, {Prandoni}, {Rafferty}, {Sabater}, {Smirnov}, {van Weeren}, {White}, {Atemkeng}, {Bester}, {Bonnassieux}, {Br{\"u}ggen}, {Brunetti}, {Chy{\.z}y}, {Cochrane}, {Conway}, {Croston}, {Danezi}, {Duncan}, {Haverkorn}, {Heald}, {Iacobelli}, {Intema}, {Jackson}, {Jamrozy}, {Jarvis}, {Lakhoo}, {Mevius}, {Miley}, {Morabito}, {Morganti}, {Nisbet}, {Orr{\'u}}, {Perkins}, {Pizzo}, {Schrijvers}, {Smith}, {Vermeulen}, {Wise}, {Alegre}, {Bacon}, {van Bemmel}, {Beswick}, {Bonafede}, {Botteon}, {Bourke}, {Brienza}, {Calistro Rivera}, {Cassano}, {Clarke}, {Conselice}, {Dettmar}, {Drabent}, {Dumba}, {Emig}, {En{\ss}lin}, {Ferrari}, {Garrett}, {G{\'e}nova-Santos}, {Goyal}, {G{\"u}rkan}, {Hale}, {Harwood}, {Heesen}, {Hoeft}, {Horellou}, {Jackson}, {Kokotanekov}, {Kondapally}, {Kunert-Bajraszewska}, {Mahatma}, {Mahony}, {Mandal}, {McKean},
  {Merloni}, {Mingo}, {Miskolczi}, {Mooney}, {Nikiel-Wroczy{\'n}ski}, {O'Sullivan}, {Quinn}, {Reich}, {Roskowi{\'n}ski}, {Rowlinson}, {Savini}, {Saxena}, {Schwarz}, {Shulevski}, {Sridhar}, {Stacey}, {Urquhart}, {van der Wiel}, {Varenius}, {Webster}, \& {Wilber}}]{Shimwell2019}
{Shimwell}, T.~W., {Tasse}, C., {Hardcastle}, M.~J., {et~al.} 2019, \aap, 622, A1

\bibitem[{{Square Kilometre Array Cosmology Science Working Group} {et~al.}(2020){Square Kilometre Array Cosmology Science Working Group}, {Bacon}, {Battye}, {Bull}, {Camera}, {Ferreira}, {Harrison}, {Parkinson}, {Pourtsidou}, {Santos}, {Wolz}, {Abdalla}, {Akrami}, {Alonso}, {Andrianomena}, {Ballardini}, {Bernal}, {Bertacca}, {Bengaly}, {Bonaldi}, {Bonvin}, {Brown}, {Chapman}, {Chen}, {Chen}, {Cunnington}, {Davis}, {Dickinson}, {Fonseca}, {Grainge}, {Harper}, {Jarvis}, {Maartens}, {Maddox}, {Padmanabhan}, {Pritchard}, {Raccanelli}, {Rivi}, {Roychowdhury}, {Sahl{\'e}n}, {Schwarz}, {Siewert}, {Viel}, {Villaescusa-Navarro}, {Xu}, {Yamauchi}, \& {Zuntz}}]{SquareKilometreArrayCosmologyScienceWorkingGroup2020}
{Square Kilometre Array Cosmology Science Working Group}, {Bacon}, D.~J., {Battye}, R.~A., {et~al.} 2020, \pasa, 37, e007

\bibitem[{{Sweijen} {et~al.}(2022){Sweijen}, {van Weeren}, {R{\"o}ttgering}, {Morabito}, {Jackson}, {Offringa}, {van der Tol}, {Veenboer}, {Oonk}, {Best}, {Bondi}, {Shimwell}, {Tasse}, \& {Thomson}}]{Sweijen2022}
{Sweijen}, F., {van Weeren}, R.~J., {R{\"o}ttgering}, H.~J.~A., {et~al.} 2022, Nature Astronomy, 6, 350

\bibitem[{{Tanaka} {et~al.}(2018){Tanaka}, {Coupon}, {Hsieh}, {Mineo}, {Nishizawa}, {Speagle}, {Furusawa}, {Miyazaki}, \& {Murayama}}]{Tanaka2018}
{Tanaka}, M., {Coupon}, J., {Hsieh}, B.-C., {et~al.} 2018, \pasj, 70, S9

\bibitem[{{Tasse} {et~al.}(2021){Tasse}, {Shimwell}, {Hardcastle}, {O'Sullivan}, {van Weeren}, {Best}, {Bester}, {Hugo}, {Smirnov}, {Sabater}, {Calistro-Rivera}, {de Gasperin}, {Morabito}, {R{\"o}ttgering}, {Williams}, {Bonato}, {Bondi}, {Botteon}, {Br{\"u}ggen}, {Brunetti}, {Chy{\.z}y}, {Garrett}, {G{\"u}rkan}, {Jarvis}, {Kondapally}, {Mandal}, {Prandoni}, {Repetti}, {Retana-Montenegro}, {Schwarz}, {Shulevski}, \& {Wiaux}}]{Tasse2021}
{Tasse}, C., {Shimwell}, T., {Hardcastle}, M.~J., {et~al.} 2021, \aap, 648, A1

\bibitem[{{The Dark Energy Survey Collaboration}(2005)}]{TheDarkEnergySurveyCollaboration2005}
{The Dark Energy Survey Collaboration}. 2005, arXiv e-prints, astro

\bibitem[{{Tunbridge} {et~al.}(2016){Tunbridge}, {Harrison}, \& {Brown}}]{Tunbridge2016}
{Tunbridge}, B., {Harrison}, I., \& {Brown}, M.~L. 2016, \mnras, 463, 3339

\bibitem[{{van Haarlem} {et~al.}(2013){van Haarlem}, {Wise}, {Gunst}, {Heald}, {McKean}, {Hessels}, {de Bruyn}, {Nijboer}, {Swinbank}, {Fallows}, \& et~al.}]{vanHaarlem2013}
{van Haarlem}, M.~P., {Wise}, M.~W., {Gunst}, A.~W., {et~al.} 2013, \aap, 556, A2

\bibitem[{Virtanen {et~al.}(2020)Virtanen, Gommers, Oliphant, Haberland, Reddy, Cournapeau, Burovski, Peterson, Weckesser, Bright, {van der Walt}, Brett, Wilson, Millman, Mayorov, Nelson, Jones, Kern, Larson, Carey, Polat, Feng, Moore, {VanderPlas}, Laxalde, Perktold, Cimrman, Henriksen, Quintero, Harris, Archibald, Ribeiro, Pedregosa, {van Mulbregt}, \& {SciPy 1.0 Contributors}}]{Virtanen2020SciPy}
Virtanen, P., Gommers, R., Oliphant, T.~E., {et~al.} 2020, Nature Methods, 17, 261

\bibitem[{{W}es {M}c{K}inney(2010)}]{McKinney2010}
{W}es {M}c{K}inney. 2010, in {P}roceedings of the 9th {P}ython in {S}cience {C}onference, ed. {S}t\'efan van~der {W}alt \& {J}arrod {M}illman, 56 -- 61

\bibitem[{{White} {et~al.}(1997){White}, {Becker}, {Helfand}, \& {Gregg}}]{White1997}
{White}, R.~L., {Becker}, R.~H., {Helfand}, D.~J., \& {Gregg}, M.~D. 1997, \apj, 475, 479

\bibitem[{Zonca {et~al.}(2019)Zonca, Singer, Lenz, Reinecke, Rosset, Hivon, \& Gorski}]{Zonca2019}
Zonca, A., Singer, L., Lenz, D., {et~al.} 2019, Journal of Open Source Software, 4, 1298

\bibitem[{{Zuntz} {et~al.}(2013){Zuntz}, {Kacprzak}, {Voigt}, {Hirsch}, {Rowe}, \& {Bridle}}]{Zuntz2013}
{Zuntz}, J., {Kacprzak}, T., {Voigt}, L., {et~al.} 2013, \mnras, 434, 1604

\end{thebibliography}

\begin{appendix}

\section{PSF leakage in HSC ELAIS-N1 data}
\label{appendix:psf_leakage}

\begin{figure}[htbp]
    \includegraphics{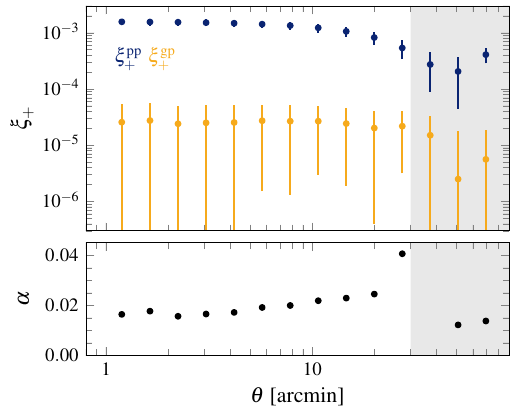}
    \caption{Estimation of the PSF leakage parameter $\alpha$. Top panel shows the cross-correlation between galaxy and PSF shapes ($\xi^\mathrm{gp}_+$), and auto-correlation of the PSF shapes ($\xi^\mathrm{pp}_+$), where the error bars are calculated from bootstrapping. The bottom panel shows the calculated $\alpha$ at each angular scale. Data points in grey region are not included in the final estimation of the mean value of $\alpha$.}
    \label{fig:alpha}
\end{figure}

To estimate the PSF systematics in the HSC shear data, we assume a linear model for the systematic introduced by the PSF leakage,
\begin{equation}
    \boldsymbol{\gamma}^\mathrm{sys} = \alpha\boldsymbol{\gamma}^\mathrm{p},
\end{equation}
where $\boldsymbol{\gamma}^\mathrm{p}$ is the ellipticity of the model PSF and $\alpha$ is a parameter that quantifies the extent of leakage. We can estimate $\alpha$ from the cross-correlation between the galaxy and PSF shapes $\xi^\mathrm{gp}_+$,
\begin{equation}
    \label{eq:alpha}
    \xi^\mathrm{gp}_+(\theta) = \alpha \xi^\mathrm{pp}_+(\theta)
\end{equation}
where $\xi^\mathrm{pp}_+$ is the auto-correlation of the PSF ellipticity. We use the galaxies from all three tomographic bins to calculate the quantities needed above. The measured $\xi^\mathrm{gp}_+$ and $\xi^\mathrm{pp}_+$ are shown in the top panel of Fig.~\ref{fig:alpha}. The bottom panel of Fig.~\ref{fig:alpha} shows the calculated $\alpha$ values as a function of scale. Since $\alpha$ is not a scale-dependent parameter, we take the average value and its standard deviation as an final estimation of $\alpha$, which leads to $\alpha=0.021\pm0.007$. Based on this parameter, the systematic from the PSF leakage can be estimated from
\begin{equation}
    \hat{\xi}_\mathrm{PSF,+}(\theta) = \alpha^2\xi^\mathrm{pp}_+(\theta).
\end{equation}

Within the angular range from 1$^{\prime}$ to 30$^{\prime}$, the average value for $\hat{\xi}_\mathrm{PSF,+}$ is $\sim$$5.7\times10^{-7}$, which is more than two orders of magnitude smaller than the detected lensing signal in Sect.~\ref{sec:hsc_2pcf}.

\section{LOFAR PSFs}
\label{appendix:lofar_psf}

\begin{figure}[htbp]
    \centering
    \includegraphics{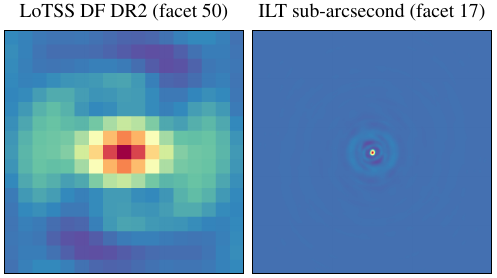}
    \caption{PSF images for ELAIS-N1 field data obtained with LoTSS Deep Fields DR2 (left), and LoTSS high resolution (right). The image size is $26^{\prime\prime}\times26^{\prime\prime}$. For colour scale, the maximum and minimum pixel values are set to $1$ and $-1$.}
    \label{fig:psf}
\end{figure}

\begin{figure}[htbp]
    \centering
    \includegraphics{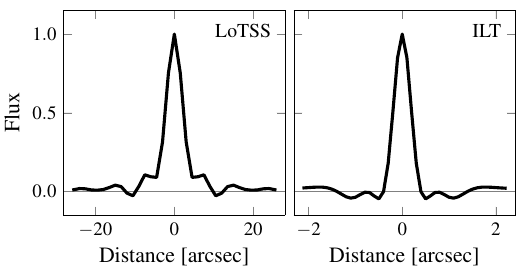}
    \caption{PSF profiles for LoTSS Deep Fields DR2 (left) and LoTSS high resolution (right). The corresponding facets of the PSFs are the same as those in Fig.~\ref{fig:psf}. Radio observations are conducted in the visibility domain, which selects a range of Fourier modes and results in the wiggling of the PSF tail around zero.}
    \label{fig:psf_profile}
 \end{figure}

For radio telescopes, the shape of the PSF on the image plane depends mainly on the radio array distribution and integrated observing time. It is in principle highly deterministic. Nonetheless, for LOFAR telescope which operates at low radio frequencies, the turbulence in the Earth's ionosphere would distort the radio emission and smear the radio image. To resolve this issue caused by ionospheric effects, a special technique called direction-dependent calibration was employed while generating the clean images in the LoTSS surveys used in this paper. This technique involves first dividing the dirty image into multiple facet images, each of which is then calibrated and deconvolved separately using its own facet PSF. In Fig.~\ref{fig:psf}, we show the PSFs from the facets that are closest to the primary beam centre. Figure~\ref{fig:psf_profile} shows the cross-sections of the LOFAR PSFs. Compared to PSFs in optical surveys, which usually can be well modelled by simple Gaussians to first order, the radio PSFs have some secondary structures caused by radio side lobes.

\begin{figure}[htbp]
    \centering
    \includegraphics{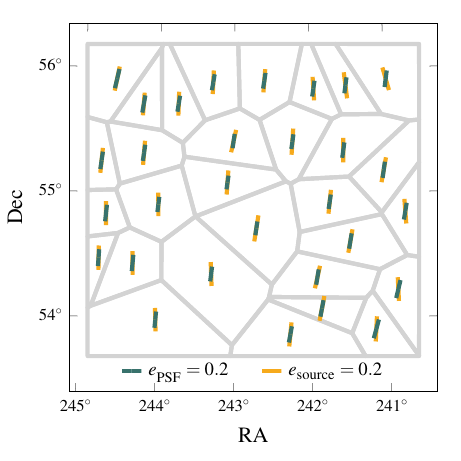}
    \caption{PSF ellipticity (green) and average source ellipticity (yellow) distributions across 30 facets in the ELAIS-N1 field from ILT observations. A clear alignment is observed between source and PSF ellipticities.}
    \label{fig:facet_psf}
 \end{figure}

There are slight variations among the PSFs across different facets, as shown in Fig.~\ref{fig:facet_psf}. While the PSF ellipticities are generally aligned in direction, small differences are evident both in orientation and ellipticity magnitude.

\section{ILT source shapes at lower resolution}
\label{appendix:ilt_weighting}

As mentioned in Sect.~\ref{subsec:trade-off}, high-resolution radio imaging can resolve out extended emission. This effect may bias the shape measurements obtained from the highest available ILT resolution of 0.3\arcsec. In this section, we examine ILT source shapes derived from the lower resolution, compare them with those obtained at 0.3\arcsec, and assess whether our choice of ILT resolution adequately preserves the shape information. We use the ILT 0.6\arcsec\ data as the lower-resolution reference. ILT 1.2\arcsec\ data are excluded from this comparison due to its lower detection sensitivity and more elongated PSF.

\begin{figure}
    \centering
    \includegraphics{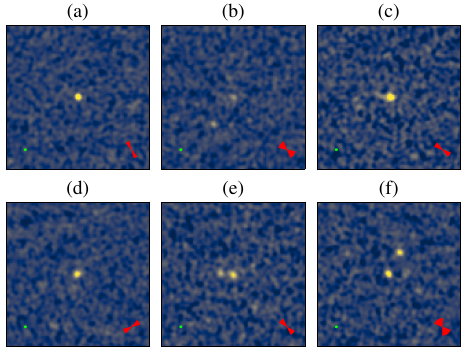}
    \caption{$21^{\prime\prime}\times21^{\prime\prime}$ cutout images of the same sources shown in Fig.~\ref{fig:cutouts}, now displayed at ILT 0.6\arcsec\ resolution. The PSF has a size of $0.58^{\prime\prime} \times 0.62^{\prime\prime}$ and is shown in the bottom-left corner of each panel. The maximum and minimum pixel values are fixed to $-3\sigma$ and $10\sigma$. The standard error of the position angle is shown in the bottom-right corner of each panel in a fan-shaped patch.}
    \label{fig:cutouts_06}
\end{figure}

We begin with a visual comparison of source shapes. Figure~\ref{fig:cutouts} in Sect.~\ref{sec:compare} presents 0.3\arcsec\ ILT cutouts for six representative sources, with corresponding 0.6\arcsec\ cutouts shown in Fig.~\ref{fig:cutouts_06}. Position angles measured at the two resolutions are generally consistent, with 0.6\arcsec values differing from those at 0.3\arcsec by 6.8$^\circ$, 25.2$^\circ$, 23.8$^\circ$, 3.7$^\circ$, 22.9$^\circ$, -0.6$^\circ$ for sources (a) to (f). Notably, source (c) exhibits more pronounced extended emission in the 0.6\arcsec\ image, with its position angle aligning more closely with that derived from the corresponding HSC image.

We further compare the total flux and position angles of \numprint{6022} sources matched between the 0.3\arcsec\ sample selected in Sect.~\ref{subsec:ILT} and the full 0.6\arcsec\ sample. Approximately 83\% of the 0.3\arcsec\ sources are also detected in the 0.6\arcsec\ low-resolution images. The undetected sources are typically too faint in the low-resolution data to be confirmed as reliable detections. We find the total flux in the 0.6\arcsec\ data is $\sim$2\% higher than in the 0.3\arcsec\ data on average. Besides, the average absolute difference in position angles is 24$^\circ$, which is smaller than the position angle uncertainties. The Pearson correlation between the position angles at two resolutions is $R_\mathrm{\cos(2\alpha)}=0.52\pm0.01$.

These results show that 0.3\arcsec\ images preserve comparable shape information from 0.6\arcsec. Using the 0.6\arcsec\ resolution would not yield significantly different results in this study.

\end{appendix}

\end{document}